\documentclass{article}
\usepackage{arxiv}
\usepackage[utf8]{inputenc} %
\usepackage[T1]{fontenc}    %
\usepackage{hyperref}       %
\usepackage{url}            %
\usepackage{booktabs}       %
\usepackage{amsfonts}       %
\usepackage{nicefrac}       %
\usepackage{microtype}      %
\usepackage{lipsum}		%
\usepackage{graphicx}
\usepackage{doi}
\usepackage{xcolor}
\usepackage{amsmath, amsfonts}

\def\mydefb#1{\expandafter\def\csname b#1\endcsname{\bm{#1}}}
\def\mydefallb#1{\ifx#1\mydefallb\else\mydefb#1\expandafter\mydefallb\fi}
\mydefallb ABCDEFGHIJKLMNOPQRSTUVWXYZabcdeghijklnopqrstuvwxyz\mydefallb
\usepackage[alignedforall]{lpform} %
\def\mydefgreek#1{\expandafter\def\csname b#1\endcsname{\text{\boldmath$\mathbf{\csname #1\endcsname}$}}}
\def\mydefallgreek#1{\ifx\mydefallgreek#1\else\mydefgreek{#1}%
	\lowercase{\mydefgreek{#1}}\expandafter\mydefallgreek\fi}
\mydefallgreek {beta}{Gamma}{alpha}{Delta}{epsilon}{Omega}{Theta}{Iota}{Lambda}{kappa}{mu}{nu}{Xi}{Pi}{chi}{rho}{Sigma}{Psi}{tau}\mydefallgreek

\usepackage{hyperref}   
\hypersetup{%
bookmarksnumbered,
pdfstartview={FitH},
linkcolor=blue,
citecolor=blue,
colorlinks=true,
pdfpagemode={UseNone},
plainpages=false
}%
\usepackage{tabularx}
\usepackage{multirow}

\usepackage{amssymb}

\usepackage{ulem}
\renewcommand{\emph}[1]{{\it #1}}
\usepackage{booktabs}
\usepackage{accents}
\newcommand{\ubar}[1]{\underaccent{\bar}{#1}}

\newcounter{condition}

\usepackage{amsthm}

\newtheorem{definition}{Definition}

\title{
Adaptive clinical trials based on design-optimal e-values with automatic curtailment:\\ An application to single-arm trials with binary data
}

\author{Stef Baas\thanks{Corresponding author} \\
    MRC Biostatistics Unit, University~of~Cambridge, East Forvie Building,\\ Forvie Site, Robinson Way, CB2 0SR, Cambridge, United Kingdom \\\phantom{.}\\ 
 {\bf Judith ter Schure}\\
     Department of Epidemiology \& Data Science, Amsterdam UMC, Amsterdam, the Netherlands.
     \\\phantom{.}\\ 
 {\bf Joost van Rosmalen}\\
     Julius Center for Health Sciences and Primary Care,\\ University Medical Center Utrecht, Utrecht University, Utrecht, the Netherlands
}
\renewcommand{\shorttitle}{Adaptive clinical trials based on design-optimal e-values with automatic curtailment}

\hypersetup{
pdftitle={Adaptive clinical trials based on design-optimal e-values with automatic curtailment},
pdfauthor={Stef Baas, Judith ter Schure, Joost van Rosmalen},
pdfkeywords={dynamic programming;  optimal betting; futility stopping; multi-stage trial; anytime-valid inference},
}

\usepackage{bm}
\def\mydefb#1{\expandafter\def\csname b#1\endcsname{\bm{#1}}}
\def\mydefallb#1{\ifx#1\mydefallb\else\mydefb#1\expandafter\mydefallb\fi}
\mydefallb ABCDEFGHIJKLMNOPQRSTUVWXYZabcdeghijklnopqrstuyz\mydefallb

\def\mydefgreek#1{\expandafter\def\csname b#1\endcsname{\text{\boldmath$\mathbf{\csname #1\endcsname}$}}}
\def\mydefallgreek#1{\ifx\mydefallgreek#1\else\mydefgreek{#1}%
	\lowercase{\mydefgreek{#1}}\expandafter\mydefallgreek\fi}
\mydefallgreek {beta}{Gamma}{alpha}{Delta}{epsilon}{Omega}{Theta}{Iota}{Lambda}{kappa}{mu}{nu}{Xi}{Pi}{chi}{rho}{Sigma}{Psi}{tau}\mydefallgreek

\usepackage{amsmath,amssymb,amsfonts}%
\usepackage{amsthm}%
\usepackage{mathrsfs}%
\usepackage[title]{appendix}%
\usepackage{textcomp}%
\usepackage{manyfoot}%
\usepackage{algorithm}%
\usepackage{algorithmicx}%
\usepackage{algpseudocode}%
\usepackage{listings}%
\algnewcommand{\Inputs}[1]{%
	\State \textbf{Inputs:}
	\Statex \hspace*{\algorithmicindent}\parbox[t]{.8\linewidth}{\raggedright #1}
}
\algnewcommand{\Initialize}[1]{%
	\State \textbf{Initialize:}
	\Statex \hspace*{\algorithmicindent}\parbox[t]{.8\linewidth}{\raggedright #1}
}
\usepackage{placeins}
\usepackage{float}
\usepackage{orcidlink}
\usepackage{hyperref}                                           
\hypersetup{%
bookmarksnumbered,
pdfstartview={FitH},
linkcolor=blue,
citecolor=blue,
colorlinks=true,
pdfpagemode={UseNone},
plainpages=false
}%

\newtheorem{theorem}{Theorem}

\newtheorem{remark}[theorem]{Remark}

\newtheorem{example}[theorem]{Example}
\usepackage[round]{natbib}
\begin{document}
\maketitle

\begin{abstract} 
The $e$-value is gaining traction as a robust alternative to p-values and Bayes factors for quantifying statistical evidence. $e$-values are a promising method for adaptive clinical trials due to their anytime-validity: $e$-values ensure type I error rate control at any stopping time, facilitating repeated interim analyses, complex stopping rules, and valid inference under protocol deviations. The $e$-value literature focuses mostly on asymptotic optimality; however, sample sizes in clinical trials are often limited.
To this end, we investigate $e$-value-based designs with finite-horizon optimality for single-arm multi-stage clinical trials with binary data. This setting is relevant in early-phase cancer trials, but it also facilitates an accessible introduction to the betting interpretation of~$e$-values, which we use to construct  $e$-values that either (1) maximize statistical power, or (2) minimize the expected sample size, with or without constraints on the minimum power. We construct these designs through (constrained) dynamic programming based on the currently observed $e$-value, the maximum sample size, and the pre-specified significance level.
Using exact calculations, we show that, next to robustness, $e$-value-based designs can provide competitive operating characteristics to standard (non-)adaptive designs with and without futility stopping and outperform growth-rate-optimal $e$-values in finite samples. In addition, small $e$-values automatically indicate trial continuation is futile, e.g., an $e$-value of zero indicates the impossibility of an efficacy conclusion.  Hence, $e$-value-based designs provide a viable alternative to the current state-of-the-art in single-arm binary trials, warranting extension to other adaptive clinical trial settings such as multi-arm multi-stage and response-adaptive designs.

\end{abstract}

\newpage 
\section{Introduction}\label{sect:introduction}
\label{s:intro}
Adaptive clinical trials provide considerable benefits in terms of the trial's operating characteristics~\citep{pallmann2018adaptive}. Adaptive elements may include  early stopping rules for futility or efficacy, sample size re-estimation, and response-adaptive randomization. These approaches may either use frequentist methods~\citep[e.g., frequentist group sequential designs, see][]{jennison2000group} or Bayesian methods~\citep{berry2010bayesian}.

While the addition of adaptive elements provides the flexibility to adapt to interim data in an ongoing trial,  standard adaptive clinical trials still have several degrees of rigidity. One degree of rigidity is that type I error rate control is usually no longer guaranteed under deviations from the trial protocol~(e.g., changes in the monitoring schedule, or adding another treatment arm). Protocol deviations can occur due to, e.g., low recruitment or event rates, dropout of the study, or emerging evidence from other related trials, and necessitate adjustments to the data analysis. Another degree of rigidity comes from the fact that once the planned maximum trial size has been reached, investigators are forced to stop the trial, and no more evidence can be gathered for treatment efficacy. Another question in adaptive designs revolves around the use of historical data to increase efficiency in the trial without type I error rate 
inflation. 

$e$-values, which are random variables expressing evidence against the null hypothesis,
 have gained increasing interest since 2019 as a robust inference method under unspecified data monitoring and optional continuation~\citep{grunwald2024}. 
As $e$-values, similar to Bayes factors, quantify evidence against a hypothesis by only looking at the trial data and not at counterfactuals, the validity of an $e$-value is more robust than standard frequentist approaches to trial protocol violations.
Unlike Bayes factors, however, $e$-values provide  frequentist error rate control, which is a minimal 
requirement for acceptance by regulatory agencies. 
In addition, $e$-values guarantee type I error rate control under optional trial continuation.
Lastly, $e$-values coming from multiple independent information sources can be combined in a valid manner~\citep{schure2022allin}, facilitating historical borrowing.
These properties make designs based on $e$-values a suitable candidate to overcome the aforementioned rigidity issues of standard adaptive designs.

We develop clinical trial designs based on~$e$-values that are optimal for a specific design with maximum sample size, termed design-optimal $e$-values.
Such $e$-values are different from standard, Growth Rate Optimal in the Worst case~(GROW), $e$-values, which are generally developed with an infinite horizon in mind and can be viewed as ``design-agnostic". 
Our approach makes $e$-values better applicable to adaptive clinical trials, 
where there is a high probability that the design realization is close to the planned design, hence optimizing trial operating characteristics given that the protocol is adhered to will lead to $e$-values that outperform standard~$e$-values in usual settings. 
We also consider the idea of the $e$-value as a principle for adaptive elements. 
In the current paper, we consider optimal futility stopping decisions based on the current $e$-value and stage of the trial, where the trial is stopped early without a conclusion if the evidence against the null hypothesis is deemed too low, but this idea extends to other adaptive elements (e.g., randomization probabilities and adaptation of the trial population). 

We consider frequentist multi-stage single-arm clinical trials with binary data. Multi-stage single-arm designs are often used for stage II cancer clinical trials and are widely applied: a recent review of adaptive clinical trials,~\citet{eltriki2024adaptive}, found that 39.1\%~(124 / 317) were single-arm trials. The widespread use of Simon's two-stage design in phase II trials~\citep{Grayling2022estimation} furthermore indicates that most single-arm trials use binary data.
In Simon's two-stage design, the trial can be stopped ad interim for futility under low success rates, and otherwise continues to the final stage for a test on efficacy. 
In addition, single-arm trials can have non-stochastic curtailment~\citep{chi2008curtailed}, where the trial is stopped when a futility or efficacy decision is certain, or stochastic curtailment~\citep{ayanlowo2007stochastically,law2019optimalcurtaileddesignssingle}, when such a decision has a high enough conditional probability. 

We compute (i) design-optimal $e$-values maximizing statistical power; (ii) design-optimal $e$-values minimizing the Expected Sample Size~(ESS) under the alternative hypothesis, and (iii) $e$-value-based designs with optimal futility stopping decisions that minimize ESS while guaranteeing a minimum power through exact computations. 
As described above, $e$-value-based designs provide more flexibility; in particular, we show that the aforementioned power constraint is satisfied even when the block sizes in a multi-stage design deviate from the plan, so long as the planned sample size is reached. 
We furthermore show that small or zero $e$-values provide automatic curtailment indicators under the maximum sample size formulation. All these properties follow logically from reasoning about optimal betting constructions; an interpretation that is central to the $e$-value literature and in this setting of discrete, and specifically binary, data, also provides a gentle introduction to anytime-valid inference for readers unfamiliar with this topic.

The paper is structured as follows: Section~\ref{Sect:model} introduces the betting analogy for~$e$-values based on binary data, introduces important concepts in the $e$-value field, and provides literature related to (design-optimal) $e$-values. Section~\ref{sect:design_optimal} introduces our $e$-value-based designs with finite-sample optimality. 
Section~\ref{sect:OC_results} provides a comparison of cumulative efficacy and futility stopping probabilities for several single-arm designs for binary data, including the novel $e$-value-based adaptive designs, based on exact computations.  Section~\ref{s:discuss}  concludes the paper and provides future research topics.
Appendix~\ref{sect:comp_policy_analysis} provides computational details and displays the optimal policies found for fully sequential designs.

\section{Single-arm binary trials, $e$-values, betting analogy, and related literature}\label{Sect:model}
In this section, we introduce the concept of $e$-values for adaptive single-arm trials with binary data, explain the betting analogy, and provide related literature to design-optimal $e$-values. A terminology table can be found in Appendix~\ref{appendix:terminology_table}.

\subsection{$e$-values for adaptive single-arm trials}\label{sect:e_vals}

The single-arm trial has binary outcomes~$Y_1, Y_2,\dots, Y_n$~(e.g., tumor response) coming from $n$ participants that are administered the treatment under investigation, where~$n$ denotes a finite trial horizon (e.g., the maximum number of patients that could be realistically recruited).  Each~$Y_i$ is assumed to be independently Bernoulli$(\theta)$  distributed with the same unknown probability~$\theta$. The goal of the trial is to test~$H_0:\theta\leq \theta_0$ for some known probability~$\theta_0$, representing an upper bound for the success rate under the current standard of care. An alternative hypothesis~$H_1:\theta\geq \theta_1$ is also specified, where~$\theta_1\geq \theta_0$ is a probability/success rate that has minimal clinical relevance, for which the treatment warrants further study, and hence~$H_0$ should be rejected with high probability/power when~$H_1$ is true. Let the probability distribution of the data under success rate~$\theta$ be denoted by~$\mathbb{P}_{\theta}$, with expectation~$\mathbb{E}_\theta.$

The $e$-value has seen major development since 2019~\citep{grunwald2024}.
We state the definition in the adaptive single-arm trial setting below.
    \begin{definition}{($e$-values)}
    An $e$-value quantifying {$\bm e$}vidence against~$H_0$ is a nonnegative random variable~$E$ such that~\begin{equation}\max_{\theta\leq \theta_0}\mathbb{E}_{\theta}[E]\leq 1.\label{eqn:e_valbd}\end{equation}
    Furthermore, a conditional $e$-value~$E_t$~(w.r.t. $Y^t$) is a function of~$Y^t$ such that~$\mathbb{E}_{\theta}[E_t\mid Y^{t-1}]\leq 1$ almost surely.
\end{definition}
By Markov's inequality, similar to~\eqref{VRineq}, we have for an $e$-value $E$ against~$H_0$ that
$$\max_{\theta\leq \theta_0}\mathbb{P}(E\geq 1/\alpha)\leq\alpha,$$
so that every $e$-value against~$H_0$ can be used as a measure of evidence against~$H_0$.
Note that conditional $e$-values against~$H_0$ can easily be combined to form new $e$-values against~$H_0$ by averaging or multiplication (under independence).

 \subsection{Betting analogy in the single arm  binary data setting}\label{sect:analogy}

In this section, we introduce the concept of an~$e$-value to design single-arm multi-stage~(e.g., phase II) clinical trial designs through a gambling analogy.

The gambling analogy works as follows:
a bettor is skeptical of~$H_0$ and thinks~$H_1$ is correct; in fact, the bettor believes that they can gain wealth by betting in favor of~$H_1$ and against~$H_0$.
Starting with a capital of~$M_0= 1$, the bettor sequentially, for every time~$t\in\{1,\dots, n\}$, bets a fraction~$B_t\in[0,1]$ of their capital~$M_{t-1}$ on the event~$\{Y_t=1\}$.
We assume that the bettor does not have foreknowledge, but bets are adaptive, meaning that~$B_t$ may be chosen as a function of (a subset of) the outcomes~$Y^{t-1}=(Y_1,\dots, Y_{t-1})$. 
In case $Y_t=1$, the bettor receives their investment back multiplied by a factor~$1/
\theta_0$, i.e.,~$M_{t}=M_{t-1}-B_tM_{t-1}+
B_{t}M_{t-1}/\theta_0$. In case $Y_t=0$ the bettor loses their investment, and~$M_{t}=M_{t-1}-B_tM_{t-1}$.
Hence, the capital of the bettor at time~$t$ can be written as:

\begin{equation}M_t=\begin{cases}
    M_{t-1}\cdot
(1-B_{t}),\quad &\text{if~$Y_t=0$,}\\
M_{t-1}\cdot
(1+B_{t}(1/\theta_0 - 1)),&\text{if~$Y_t=1$,}
\end{cases}\implies M_t=\prod_{t'=1}^t
(1+B_{t'}(Y_{t'}/\theta_0-1))
.\label{capital}\end{equation}

From the above, we see that the bettor can split their capital and place a fractional bet, allowing them to invest any fraction of their current capital. This betting analogy has a clear interpretation for binary outcomes: when the bet is positive, the bettor is betting that the treatment will be a success for the next participant. If the bettor believes the game is unfavorable and the capital will go down in expectation, it will be best to not place bets, i.e.,~$B_t=0$. 

Because the capital $M_t$ is bounded for every fixed~$t,$~$M_0=1$ and~$\mathbb{E}_{\theta}[M_{t+1}\mid Y^{t}]\leq \mathbb{E}_{\theta}[M_{t}\mid Y^{t}]$ under~$H_0$, we have that~$(M_t)_t$ is a {\it test supermartingale}~\citep{shafer2011testmartingales}, which is essentially a sequence of random variables starting at~$1$ and which is not expected to increase at each time~$t$ given the data~$Y^t$. 
The probability that the capital exceeds a given value is at most inversely proportional to that value by the Ville-Robbins inequality~\citep[see, e.g.,][Corollary~1]{grunwald2024}. In particular, for~$M^*>0$, \begin{equation}\max_{\theta\leq\theta_0}\mathbb{P}_{\theta}\left(\max_{t\leq n} M_t\geq M^*\right)\leq1/M^*.\label{VRineq}\end{equation}
 Alternatively,~\eqref{VRineq} indicates that one can reject the hypothesis~$H_0$ with a frequentist type I error rate bounded by~$\alpha$ at the first time~$\tau_{1/\alpha}$ for which the bettor's capital~$M_t$ has surpassed a level~$M^*=1/\alpha.$
By  the optional stopping theorem, we see that~$\mathbb{E}_{\theta}[M_{\tau}]\leq 1$ for every stopping time~$\tau$, hence $M_{\tau}$ and in particular~$M_{\tau_{1/\alpha}}$ is an~$e$-value for every stopping time~$\tau$.
The capital process~$(M_t)_t$ can hence also be referred to as an~$e$-process~\citep{grunwald2024}, which is the term we will be using from now on, as this is more in line with reality: we are not really betting on outcomes, but we are collecting evidence against~$H_0.$
Under the null hypothesis, we hence see that the expectation of the capital is 1 or smaller, making the bet fair for ``the house'' under the null hypothesis~\citep[it expresses no expected increase in the bettor's capital,][]{ramdas2023game}. Inverting this statement, if the bettor's capital were to ever become large, this would count as evidence against~$H_0$.

In single-arm trials with binary data, we can encounter situations that render further recruitment pointless, as any possible realization of the remaining data in the pipeline will lead to a non-rejection with certainty.  A concrete example is a fifty-patient trial that aims to reject the null hypothesis of a 30\% proportion of patients with a good outcome. A right-sided test with a significance level of 5\% would need to observe a proportion of at least 40\%, or 20 out of 50 patients with the good outcome. If we only observe 9 successes in the first 40 patients, it is impossible to get that number up to 20 with the remaining 10 patients. A similar phenomenon happens in the betting analogy when the capital or evidence against~$H_0$ becomes zero before the final analysis point: without any money to bet with, bets cannot continue. \citet{koning2026anytime} developed a method to construct an anytime-valid test from a finite sample test through the conditional rejection rate under~$H_0$ divided by~$\alpha.$ Using this transformation, we see that non-stochastic curtailment for futility always corresponds to evidence against~$H_0$ becoming equal to zero. Similarly, non-stochastic curtailment for efficacy, i.e., stopping when an efficacy conclusion is certain, corresponds to the evidence against~$H_0 $ reaching~$1/\alpha$ before the final analysis point, as the conditional rejection rate under~$H_0$ becomes one.

In the betting analogy introduced above, the bets~$B_t$ are unspecified apart from the requirement~$B_t\in[0,1]$ for all~$t$. The next section gives the most common choice.

\subsection{Kelly betting and $e$-values that are Growth Rate Optimal in the Worst-case (GROW)}\label{sect:GROW}
\citet{grunwald2024} defined the Growth Rate Optimal in Worst-case~(GROW) criterion to select $e$-values based on their minimum growth rate across parameter values under~$H_1$. In the betting analogy, if~$M_t$ equals the GROW $e$-value for all~$t$, this means that the asymptotic growth rate for the $e$-value is maximized at the boundary~$\theta=\theta_1$, i.e., in the betting analogy we are placing Kelly bets~\citep{kelly_new_1956}. We state the GROW property here for convenience:

\begin{definition}[Growth Rate Optimality in Worst-case~(GROW)]~\label{def:GROW}
    Let~$\mathcal{E}_0$ be the set of $e$-values against hypothesis~$H_0$.
    An $e$-value~$E$ is growth rate optimal in worst-case under hypothesis~$H_1$ if 
    $$E=\underset{E'\in\mathcal{E}_0}{\arg\max}\min_{\theta\geq\theta_1}\mathbb{E}_{\theta}[\log(E')].$$
\end{definition}
The GROW criterion maximizes the log-$e$-value, hence, GROW $e$-values  cannot become zero with nonzero probability~(i.e.,~$\mathbb{P}_{\theta_1}(E=0)=0$ for all~$\theta_1$). So GROW $e$-values never yield non-stochastic curtailment; when optimizing for an infinite horizon, we want to ensure we can continue betting indefinitely. 
Theorem 1 in \cite{grunwald2024} implies that $(M_t)_t$ is a GROW $e$-process, meaning all multiplication terms in~\eqref{capital} are GROW $e$-values, when setting each~$B_t$ equal to the Kelly bet \begin{equation}B_{\text{Kelly}}=\max_{b\in[0,1]}\mathbb{E}_{
\theta_1
}[\log(1-b(Y_{1}/\theta_0-1))]=(\theta_1-\theta_0)/(1-\theta_0).\label{eq:Kellybet}\end{equation}  
From~\eqref{capital}, letting~$S_t=\sum_{t'=1}^tY_{t'}$ it follows that the GROW $e$-process equals
\begin{equation*}
    M_t^{\text{GROW}}=\prod_{t'=1}^t\left(\frac{\theta_1}{\theta_0}\right)^{Y_{t'}}\left(\frac{1-\theta_1}{1-\theta_0}\right)^{1-Y_{t'}} = \left(\frac{\theta_1}{\theta_0}\right)^{S_{t}}\left(\frac{1-\theta_1}{1-\theta_0}\right)^{t-S_{t}},
\end{equation*}
hence, $M_t^{\text{GROW}}$ equals the sequential likelihood ratio~\citep{wald1945sequential} for~$\theta=\theta_0$ vs. $\theta=\theta_1$.
As shown in~\citet{breimualphanu2011optimal}, $M^{\text{GROW}}=(M_t^{\text{GROW}})_t$ is the $e$-process that asymptotically reaches a threshold~$M^*$ the quickest when $\theta=\theta_1$, i.e., $\lim_{M^*\rightarrow\infty}\mathbb{E}_{\theta_1}\left[\tau_{M^*}^{M'} - \tau_{M^*}^{M^\text{GROW}} \right]\geq0$ for all $e$-processes~$M'$, where~$\tau_{M^*}^{M}=\inf\{t:M_t\geq M^*\}$.
In addition, any other $e$-process
is asymptotically dominated by the GROW $e$-process, i.e.~$\mathbb{E}_{\theta_1}\left[\lim_{t\rightarrow\infty}M'_t/M^{\text{GROW}}_t\right]\leq 1$ for all $e$-processes~$M'$. Both limits above are not representative of clinical trial settings, as the significance level~$\alpha = 1/M^*$ is usually fixed~(in contrast to~$M^*\rightarrow\infty$) and there is often a finite limit on the maximum possible trial size, e.g., in rare disease settings~(in contrast to~$t\rightarrow\infty$).

\subsection{Related literature}

First of all, the $e$-process~\eqref{capital} has two well-known alternative formulations
in the literature:

\begin{itemize}
\item{\bf Ticket formulation} \citep{WaudbySmithbettingmeans}\label{alt_capital}\\
The formulation~$(V_t)_t$ below is a general $e$-process testing the mean for bounded random variables~$Y_t\in[0,1]$:
    \begin{align*}V_t&=\prod_{t'=1}^t
(1+\lambda_{t'}(Y_{t'}-\mu_0)).\end{align*}
In the above,~$\lambda_t\in[-1/(1-\mu_0),1/\mu_0]$ to ensure each product-term is nonnegative.

The process~$(V_t)_t$ reduces to~$(M_t)_t$ for binary data with $\mu_0 = \theta_0$ when~$\lambda_t\geq 0$ and~$\lambda_t=B_t/\theta_0$. We can only directly relate~$V_t$ to~$M_t$ if~$\lambda_t\geq0$ because we focus on one-sided instead of two-sided testing, i.e., we are only interested in finding evidence against~$\theta\leq\theta_0$ and not~$\theta=\theta_0$. We call the process~$(V_t)_t$ the ``ticket formulation'' as it is analogous to paying for~$\lambda_t$ tickets with price~$\mu_0$ that yield a reward~$Y_t.$
We present the results of this paper in the representation $M_t$ instead of $V_t$ because it is easier to interpret the betting fraction $B_t$ in the range~$[0,1]$, which does not depend on the null hypothesis boundary $\theta_0$.
\item{\bf Sequential likelihood ratio formulation} \citep{grunwald2024}\label{alt_capital}\\
The formulation~$(W_t)_t$ below is a sequential binary data Bernoulli likelihood ratio with~$Q_t\in[0,1]$:
    \begin{align*}
    W_t &=\prod_{t'=1}^t\left(\frac{Q_{t'}}{\theta_0}\right)^{Y_{t'}}\left(\frac{1-Q_{t'}}{1-\theta_0}\right)^{1-Y_{t'}}.\end{align*}\\
Instead of betting a fraction $B_t$ of the capital only on the event ${Y_t = 1}$, and explicitly saving a fraction $(1-B_t)$, this formulation expresses the same investment by betting on both $Y_t = 1$ and $Y_t = 0$ and splitting the entire capital between them, based on fractions $Q_t$ and $1-Q_t$~(i.e., explicit hedging). Since either of them will pay out if the bet on both is non-zero, the bettor saves a non-zero fraction of their capital, just as for non-zero~$(1-B_t)$. We can relate~$W$ to~$M$ as $W_t= M_t$ for all~$t$ if~$Q_t=\theta_0+(1-\theta_0)B_t$:
\begin{equation*}
    W_t =W_{t-1}\cdot\begin{cases}
    \frac{\theta_0+(1-\theta_0)B_t}{\theta_0}= 1+ B_t(1/\theta_0-1),\quad&\text{if~~$Y_t=1$,}\\
      \frac{1-\theta_0-(1-\theta_0)B_t}{1-\theta_0}= 1-B_t,&\text{if~~$Y_t=0$.}\end{cases}
\end{equation*}
\citet{shafer2021testingbetting} calls $Q_t\in[\theta_0,1]$ the alternative (parameter) implied by the bet~$B_t$, as it equals the numerator in the likelihood ratio parameterization. We present the results of this paper in the representation $M_t$ instead of $W_t$ because the range for $B_t$ from $0$ to $1$ is easier to interpret as betting less or more aggressive, with the capital at risk ranging from nothing to everything. 
\end{itemize}

The concept of a (test) martingale has been around since 1939, and was introduced in the thesis of Jean Ville~\citep{ville1939etude}. A further discussion on the historical use of test martingales and its relation to standard tests, including sequential analysis as all-or-nothing tests, introduced by Abraham Wald and George A. Barnard, can be found in~\citet{shafer2011testmartingales}. Since 2019, there has been an increase in interest in anytime-valid inference~\citep{grunwald2024}, with an introduction and overview of recent advances in the field given by~\citet{ramdas2023game,ramdas2025hypothesis}.

More recently, several papers have come out that focused on the finite-sample performance of $e$-values. \citet{voravcek2025star} considered the construction of anytime-valid confidence sequences that, for finite horizons, improve in radius over the anytime-valid confidence sequences of~\citet{WaudbySmithbettingmeans}. 
They propose two algorithms that explicitly take into account how many rounds the bettor still has, as well as the distance from the threshold. Their first algorithm, STaR, shows improvement in width over standard Hoeffding and Bernstein bounds, while their second algorithm, STaR-Bets, is within a factor $(1+o(1))$ of the optimal finite-sample confidence width. In contrast to our work, \citet{voravcek2025star} do not consider dynamic programming to find optimal bets. 

 \citet{taga2026learning} developed  horizon-aware anytime-valid tests
and confidence sequences for bounded outcomes through an optimal control problem, maximizing statistical power at a fixed horizon. They characterize the behavior of the optimal betting strategy, showing that betting more aggressively than Kelly betting is better if the bettor is behind, and betting less aggressively than Kelly betting is better when ahead.
In the general setting, they approximate the optimal betting strategy using a reinforcement learning approach, yielding state-of-the-art results. In contrast to our work, \citet{taga2026learning} only consider maximizing power and furthermore only consider three bet sizes~(conservative, Kelly, aggressive).

 \citet{ clerico2026time}
consider a testing-by-betting framework using an optimal control problem with rewards favoring early rejection. 
They show for the simple null versus simple alternative case that the~$e$-value maximizing power in finite-horizons, is induced by a Neyman Pearson test on the outcome paths. Unlike our betting strategies, the resulting optimal betting strategy for binary outcomes is, however, non-Markovian.  
For softer deadlines, they introduce the EDO criterion
for exponentially decaying rewards, which is a
finite-time counterpart to growth-rate-optimality. In contrast to our work, \citet{clerico2026time} only considers maximizing power in the finite horizon setting.

\citet{koning2026anytime} provide an explicit bridge between classical testing using finite horizons and anytime validity, and provide a method to ``sequentialize a test''  using a Doob martingale. 
In addition, they provide a method for optional continuation under standard tests. 
In Section~6, they provide a discussion relating GROW $e$-values to Neyman-Pearson tests. Similar to our approach and the approach in \citet{koyamaproper2008}, the $e$-value construction of~\citet{koning2026anytime} can be used to redesign the trial with a new significance level at each protocol deviation. However, our approach leads to stronger optimality guarantees, and (for instance) guarantees a minimum power even under deviations from the planned stage sizes.
 
 The critical choice of the bet size for the infinite horizon betting problem has been explored in many papers, and an overview is given in Section~2 of~\citet{taga2026learning}.  In addition, \citet{fisher_ramdas2026} developed a sequential boosting method to improve test supermartingales, in the sense that they lead to earlier rejection, by avoiding overshoot at level~$1/\alpha$. In this paper, we focus explicitly on finite-horizon performance, but find that our design-optimal $e$-values also avoid overshoot. We note this is mostly due to our tie-breaking rule in the optimization instead of by necessity.

\section{$e$-value-based designs with finite-time optimality}\label{sect:design_optimal}

\subsection{Design-optimal $e$-values}\label{sect:DO_e}
This section introduces design-optimal $e$-values which are based on an optimal finite-horizon betting strategy that takes the {\it trial design} into account. This is different from GROW $e$-values, which have asymptotic optimality guarantees.
The rationale of the design-optimal $e$-values we consider is to bet more aggressively in case the $e$-value still needs a large increase to reach the rejection boundary $1/\alpha$ when close to the maximum sample size, i.e., when the bettor is ``behind schedule'', and to bet less aggressively when the $e$-value is large and there are many more observations before reaching the maximum sample size, i.e., if the bettor is ``ahead''~\citep{taga2026learning}. We give two examples.

\begin{example}[Going all-or-nothing]\label{example:all_nothing}
    Assume that we have observed binary data $Y_1, Y_2, \ldots, Y_{n-1}$, resulting in~$M_{n-1}=10$ and that the bettor can at most double their capital if $Y_n = 1$, i.e.,~$\theta_0=1/2$. Then the only way in which they can reach $M_n\geq 20$, rejecting~$H_0$ with significance level $\alpha = 0.05$, is by betting everything on $Y_n = 1$. If we observe $Y_n = 1$, we reject~$H_0$. If we observe $Y_n = 0$, the bettor has lost all capital and is left with $M_n = 0$.
    So our final bet is the maximally aggressive all-or-nothing bet. This bet can be related to using a standard test at the final analysis point, which will either lead to a rejection~(evidence $M_n=20$) or non-rejection without option for continuation~($M_n=0$)~\citep[see also][]{shafer2011testmartingales}.
\end{example}

\begin{example}[Curtailment by bankruptcy] \label{example:bankruptcy}Assume that we have observed binary data $Y_1, Y_2, \ldots, Y_{n-3}$, $M_{n-3}=0.02$ and that the bettor can at most increase their capital tenfold, i.e.,~$\theta_0=1/10$. Then the only way in which $M_n\geq  20$ can be attained and~$H_0$ can be rejected at level~$\alpha=0.05$, is when the bettor bets everything on $Y_{n-2} =Y_{n-1}=Y_{n}= 1$. If we observe $Y_{n-2}=0$, the bettor has no more capital and cannot continue. As the $e$-process against~$H_0$ cannot reach the efficacy boundary, the trial can be curtailed for futility. The link between conditional rejection rate and~$e$-values in~\citet{koning2026anytime} effectively shows that standard non-stochastic curtailment in single-arm designs directly corresponds to the $e$-value of the sequentialized binomial test reaching zero.
\end{example}

Our design-optimal $e$-values are found through a finite-horizon {\it Markov decision process}~\citep[MDP,][]{puterman1994markov}. 
An MDP consists of states, actions, transition probabilities, and rewards, which are described below. For an analysis time~$t\in[n]=\{0,\dots, n\}$, representing the number of outcomes that have been analyzed, let~$\bX_t=(t,M_{t})$ be the {\it $e$-state} of the trial, where~$\bX_0=()$. The~$e$-state encodes the degree of evidence against~$H_0$ at time~$t$, and this state can then be used to guide decision making~(e.g., bets) for subsequent stages of the trial. The state space may be truncated by assuming a significance level~$\alpha$ for which~$H_0$ will be rejected, and hence the trial will be stopped for efficacy, under the event in~\eqref{VRineq}, which means that we can restrict the state space to~$\mathcal{X}=[n]\times[0,1/\alpha].$

\begin{remark}[Hopeless zone and almost hopeless zone]\label{remark:hopeless_almost}
    From examples~\ref{example:all_nothing} and~\ref{example:bankruptcy} we note that, if at any point~$t$ during the trial we have that~$M_t< \theta_0^{n-t}/\alpha$, then it is no longer possible for the $e$-process to hit the boundary~$1/\alpha$. We refer to this subset of~$\mathcal{X}$ as the ``hopeless zone''~\citep[HZ,][]{taga2026learning}, in which it might be better to stop the trial for futility as an efficacy decision is impossible. In addition, when $\theta_0^{n-t}/\alpha<M_t< \theta_0^{n-t-1}/\alpha$, the trial is in the ``almost hopeless zone''~(AHZ) where one treatment failure before the end of the trial will lead to reaching the hopeless zone.
\end{remark}

As {\it action} in state~$\bX_t$, the bettor selects a bet size~$B_{t+1}$ for outcome~$t+1$. 
Following MDP notation, we let~$\pi$ be the betting strategy, a function mapping the ($e$-)state to the action~(bet), i.e.,~$\pi(\bX_t)=B_{t+1}.$
 Let~$m$ be a function such that~$m(\bX_t)=M_{t}$. Assuming~$m(\bX_t)<1/\alpha$, after choosing a bet size~$B_{t+1}\in[0,1]$ in state~$\bX_t$, the state follows the following update rule, where~$(Y_t)_t\stackrel{iid}{\sim} \text{Ber}(\theta_1)$:
\begin{equation}\bX_t=(t,M_t)\stackrel{B_{t+1}}{\rightarrow}\bX_{t+1}=
    (t+1,\,\min(1/\alpha, m(\bX_t)\cdot (1-B_{t+1}(Y_{t+1}/\theta_0-1))).\label{eq:transition}
\end{equation}
Furthermore, if~$M_t=1/\alpha$, then $\bX_t=(t,1/\alpha)\stackrel{B_{t+1}}{\rightarrow}\bX_{t+1}=
    (t+1,1/\alpha)$. 
Note that~$B_{t+1}=0$ yields~$M_{t+1}=M_{t}$ and~$B_{t+1}=1$ yields a probability~$1-\theta_1$ that~$M_{t+1}=0$, which is an absorbing state, after which~$M_{t+1}=M_{t+2}=\cdots=M_n=0.$
    
Hence,~$(\bX_t)_t$ describes the evolution of the $e$-process in the setting~$\theta=\theta_1$, and we denote the expectation over paths~$(\bX_t)_t$ under a strategy~$\pi$ by~$\mathbb{E}^\pi_{\theta_1}.$
In order to optimize the bets, we introduce reward functions~$r(\bX_t, B_{t+1})$, representing the reward of placing bet~$B_{t+1}\in [0,1]$ in state~$\bX_t\in \mathcal{X}$, and define the optimization problem
\begin{align}
    \max_{\pi} \mathbb{E}^\pi_{\theta_1}\left[\sum_{t=0}^nr(\bX_t,\pi(\bX_t))\right].\label{opt_prob}
\end{align}

Under conditions stated in~\citet{puterman1994markov}, the optimal solution~$\pi^*$ of~\eqref{opt_prob} can be found recursively by solving the Bellman equations for all states~$\bX_t$:
\begin{align}
    \pi^*(\bx_n)&=\underset{b_{n+1}\in[0,1]}{\arg\max}\;r(\bx_n,b_{n+1}),\label{Beq1}\\ \pi^*(\bx_t)&=\underset{b_{t+1}\in[0,1]}{\arg\max}\;\left(r(\bx_t,b_{t+1}) + \mathbb{E}_{\theta_1}[V_{\theta_1}^{\pi^*}(\bX_{t+1})\mid \bX_t=\bx_t,B_{t+1}=b_{t+1}]\right).\label{Beq2}
\end{align}
In the above,~$V_{\theta_1}^{\pi}(\bX_{t+1}) = \mathbb{E}^{\pi}_{\theta_1}\left[\sum_{t'=t+1}^{n}r(\bX_{t'},\pi(\bX_{t'}))\mid \bX_{t+1}\right]$ is the expected sum of rewards under betting strategy~$\pi$ from the next interim until the end of the trial. In~\eqref{Beq1} and~\eqref{Beq2}, ties in the maximization are broken by choosing the smallest bet size leading to the highest value.
 
In general, reward function~$r$ can be chosen as any function of the state~$\bX_t$ and bet-size~$B_{t+1}$.
We consider two choices of rewards, resulting in the power-maximizing and expected sample size-minimizing design-optimal $e$-values.

\subsubsection{Power-maximizing $e$-value}
The power-maximizing design-optimal~(P-max) $e$-value maximizes the power at the final analysis~$n$. The rewards assume the following form, where~$t<n$:~$$r(\bX_t,B_{t+1})=0,\quad r(\bX_n,B_{n+1})=\mathbb{I}(m(\bX_n)\geq 1/\alpha).$$
Hence, assuming~$m(\bX_t)<1/\alpha$, as~$M_t=1/\alpha$ is an absorbing state,~\eqref{Beq1} and~\eqref{Beq2} indicate that the optimal bets at stage~$t$ solve
$$\pi^*(\bx_t)=\underset{b_{t+1}\in[0,1]}{\arg\max} \;\mathbb{P}_{\theta_1}^{\pi^*}\left(\max_{t'\in\{t+1,\dots, n\}}M_{t'}\geq 1/\alpha\mid \bX_t=\bx_t, B_{t+1}= b_{t+1}\right).$$
For the final analysis~$n$, assuming~$M_{n-1}<1/\alpha$, there is effectively only one remaining bet, $B_n$, which maximizes~\begin{align}&\mathbb{P}_{\theta_1}\left(M_{n-1}(1-B_{n}(Y_n/\theta_0-1))\geq 1/\alpha\mid \bX_{n-1}\right)=\mathbb{P}_{\theta_1}\left(\log(1-B_{n}(Y_n/\theta_0-1))\geq \log(1/(\alpha M_{n-1})\mid \bX_{n-1}\right)\nonumber\\&=\mathbb{P}_{\theta_1}\left(Y_n\geq \left(\log(1/(\alpha M_{n-1})) - \log\left({1-B_n}\right)\right)/\log\left(\frac{1+B_n(1/\theta_0-1)}{(1-B_n)}\right)\mid \bX_{n-1}\right)\label{eq:powervalue}.\end{align}
Hence, $B_{n}$ minimizes
\begin{equation}\left(\log(1/(\alpha M_{n-1})) - \log\left({1-B_n}\right)\right)/\log\left(\frac{1+B_n(1/\theta_0-1)}{(1-B_n)}\right),\label{eq:probterm}\end{equation}
which does not depend on~$\theta_1.$
Letting~$Q_n(B_n)=\theta_0 + (1-\theta_0)B_n$ be the implied alternative by bet~$B_n$, it can be verified that the derivative of the above term is zero whenever
\begin{equation}
    r_n(B_n)=\mathbb{E}_{Q_n(B_n)}[\log(1-B_{n}(Y_n/\theta_0-1))]= \log(1/(\alpha M_{n-1})).\label{match_rate}
\end{equation}
If~\eqref{match_rate} holds, we see that~$$M_{n-1}e^{ r_n(B_n)} = M_{n-1}/(\alpha M_{n-1})=1/\alpha.$$
As~$r_n(B_n)$ equals the expected growth rate of the $e$-process under the implied alternative~$Q_{n}(B_n)$, we conclude that it is power-optimal to choose the bet~$B_n$ and implied alternative $Q_{n}$ so that the expected growth rate matches the necessary growth rate  for~$M_n$ to equal the threshold~$1/\alpha$ in stage~$n$ given~$M_{n-1}$.
From~\eqref{match_rate} we also see that the optimal bet size reduces overshoot, as the optimal bet size is expected to make the~$e$-value match the level~$1/\alpha$ exactly at the end of the trial. 
As~$Q_{n}\geq \theta_0$, we see that lower values of~$M_{n-1}$ correspond to higher optimal values of~$B_{n}$ and vice versa.
We see from~\citet[][Equation (18)]{johnson2013uniformly} that the final conditional~$e$-value equals the {\it uniformly most powerful Bayesian test} for a binomial success probability. 
Note that~$\min_{b\in[0,1]}r_n(b)=0$ and~$\max_{b\in [0,1]}r_n(b)=\log(1/\theta_0)$ whereas~$\log(1/\alpha M_{n-1})\in (0,\infty).$ If~$M_{n-1}/\theta_0\leq 1/\alpha$ the boundary can no longer be reached after one bet and~\eqref{match_rate} has no solution, as the power is equal to zero, there is no unique optimal bet~$B_n$ in this setting. While the optimal bet~$B_{n}$ does not depend on~$\theta_1$, the bets coming before the final interim depend on~$\theta_1$ through equations~\eqref{Beq1} and~\eqref{Beq2}.

\begin{example}[sub-optimality of Kelly bets]\label{example:notgrow}
     Assume we are at stage~$n-1$ with~$Y_1,Y_2,\dots, Y_{n-1}$ observed so far, with an~$e$-value of~$12$, that can be doubled if $Y_n = 1$, i.e., $\theta_0=1/2$, and that~$\theta_1=0.8$. According to~\eqref{match_rate} the probability of reaching~$1/\alpha=20$~(for~$\alpha=0.05$) is maximized when~$r(B_n)=\log(20/12)$ implying an optimal bet~$B_n=B_n^*\approx0.91$ whereas the Kelly/GROW bet~$B_{\text{Kelly}}$ in~\eqref{eq:Kellybet} equals~$0.6$. When using the bet~$B_n^*$,~\eqref{eq:probterm} is smaller than~1 yielding a conditional power of~$\theta_1=0.8$, whereas using~$B_{\text{Kelly}}$,~\eqref{eq:probterm} is larger than~1, yielding a conditional power of zero. 
\end{example}

\begin{remark}[Discreteness of outcomes]
Since~$Y_n\in\{0,1\}$ the conditional power in~\eqref{eq:powervalue}
    can essentially only attain three numbers,~$0,\theta_1,$ or~$1$. For~$M_{n-1}< 1/\alpha$, the expression in~\eqref{eq:probterm} is always larger than zero and hence the power is never equal to 1, leaving~$0$ and~$\theta_1$ as options. The conditional power goes from 0 to~$\theta_1$ at the boundary
    $$\ubar{B} = (\theta_0/(\alpha M_{n-1})-\theta_0)/(1-\theta_0) $$
    and in Example~\ref{example:notgrow} we saw that~$B_{\text{Kelly}}=0.6$ was below this boundary. Essentially every bet between~$\ubar{B} $ and~$1$ yields the same conditional power; the solution~$B_n^*$ to~\eqref{match_rate} is one option. We note that any bet in the region~$(\ubar{B},B_n^*)$ could be preferable to~$B_n^*$ as more capital is saved while attaining the same power. Our numerical approach~(Section~\ref{sect:comp_policy_analysis}) will automatically choose bets in the region $(\ubar{B},B_n^*)$ by discretizing the bet sizes and breaking ties by choosing the smallest bet size.
\end{remark}

\subsubsection{Expected sample size minimizing $e$-value}\label{sect:ESS_min_e}
Next to the P-max $e$-value, we consider a design-optimal $e$-value minimizing the expected sample size~(ESS-min $e$-value) under~$\theta=\theta_1$. This is similar to the two-stage design in~\citet{MANDER2010572}, minimizing the expected sample size under the alternative.  The rewards assume the following form for all~$t\leq n$:
\begin{equation}
    r(\bX_t,B_{t+1})=-\mathbb{I}(m(\bX_t)<1/\alpha),\label{eqn:rewards_ESS_min_e}
\end{equation}
i.e., for every time~$t<n$ where~$M_{t}<1/\alpha$ we add a penalty~$-1$ for recruiting participant~$t+1$. The final cost $-\mathbb{I}(m(\bX_n)< 1/\alpha)$  gives an incentive for the capital to grow above the level~$1/\alpha$ at the end of the trial and effectively determines the optimal bets in the penultimate trial stage. Defining the minimum of the empty set to equal infinity, let~$\tau_{1/\alpha}=\min\{t\in [n]:m(\bX_t)\geq 1/\alpha\}$ be the random time that the $e$-value first hits the level~$1/\alpha$. Assuming~$m(\bX_t)<1/\alpha$,~\eqref{Beq1} and~\eqref{Beq2} indicate that the optimal bets at stage~$n$ solve
\begin{equation}\pi^*(\bx_t)=\underset{b_{t+1}\in[0,1]}{\arg\min} \;\mathbb{E}^{\pi^*}_{\theta_1}\left[\min(n+1,\tau_{1/\alpha})-t\mid \bX_t=\bx_t, B_{t+1}= b_{t+1}\right].\label{eq:cost_ESSmin}\end{equation}
Due to the final cost, the final bet for the ESS-min $e$-value agrees with the final bet for the P-max $e$-value defined in the last section. We note that, due to the stage-wise cost~$-\mathbb{I}(m(\bX_t)<1/\alpha)$, every bet size~$B_{t+1}>0$ yields a lower cost~(the term to be minimized in~\eqref{eq:cost_ESSmin}) than the bet size~$B_{t+1}=0$ which would deterministically yield~$M_{t}=M_{t+1}.$ Hence, a bet-size $B_{t+1}=0$ will not be chosen unless~$\mathbb{P}_{\theta_1}(\tau_{1/\alpha}\leq n\mid \bX_t, B_{t+1}=b_{t+1})=0$ for all~$b_{t+1}\in[0,1]$. Furthermore, we see from~\eqref{eq:cost_ESSmin} that reaching the absorbing state~$M_t=0$ yields a deterministic cost~$n+1-t$.

    \subsection{$e$-value-based adaptive designs: Adding adaptive elements based on $e$-values}\label{sect:integrating_eval_design}

    This section illustrates the concept of integrating adaptive design elements with~$e$-values in a relatively simple setting. Namely, we introduce designs with optimal {\it futility stopping rules} based on~$e$-values.

   In trials with an analysis framework based on $e$-values, it makes sense to define stopping rules based on the $e$-value.  For example, if the $e$-value drops below a prespecified value, the trial may be stopped as the probability is low that the $e$-value will reach the efficacy boundary before the end of the trial, i.e., we propose stochastic curtailment~\citep{law2019optimalcurtaileddesignssingle} based on the~$e$-value.    As these stopping rules do not come with a rejection of the null or alternative hypothesis, they are inherently different from the usual stopping boundaries constructed using $e$-values, since $e$-values quantify the evidence against a hypothesis. 

   Futility stopping based on the~$e$-value may be done by calibrating futility boundaries based on the~$e$-value, calibrating stopping rules based on the conditional power under the~$e$-value and defining a ``hopeless zone''~\citep{taga2026learning}, but in this paper, we advocate for optimal stopping decisions integrated in the MDP underlying the design-optimal $e$-value, while ensuring a high enough overall power of the study.

   Specifically, first, we augment the action space for the MDP defined in Section~\ref{sect:DO_e} with the option to stop the trial, denoted by~$B_t=\emptyset$. For $B_t\in[0,1]$ and~$t<n$, we consider the rewards in~\eqref{eqn:rewards_ESS_min_e} with the adjustment~$r(\bX_n,\emptyset)=r(\bX_n, B_{n+1})=r((t, 0), B_{t+1})=0$ for all~$B_n$, i.e., recruiting stops when the futility stopping decision is made, at the end of the trial, or when the capital runs out. Second, we add a constraint to Optimization Problem~\eqref{opt_prob} that ensures high enough power. Particularly, letting~$\tau(\pi)=\min\{t\in [n]:m(\bX_t)\geq 1/\alpha\text{ or }\pi(\bX_t)=\emptyset\text{ or }m(\bX_t)=0\}$ be the time at which the trial stops for futility or efficacy, for a maximum type II error~$\beta\in[0,1]$, we solve the following optimization problem:
   \begin{align}&\min_{\pi}\mathbb{E}^{\pi}_{\theta_1}\left[\min(n,\tau(\pi))\right]\nonumber\\&\text{s.t.}\nonumber\\
   &\mathbb{P}_{\theta_1}^{\pi}\left(\max_{t\in[n]}M_{t}\geq 1/\alpha\right)\geq 1-\beta.\label{ineq:power_constraint}
   \end{align}
The Power Constraint~\eqref{ineq:power_constraint} removes the requirement of the final cost~$r(\bX_n, B_{n+1})=-\mathbb{I}(m(\bX_n)<1/\alpha)$ in~\eqref{eqn:rewards_ESS_min_e}; without adding this constraint, the optimal action would be to stop for futility in every state. 

Appendix~\ref{sect:comp_policy_analysis} provides computational details for the design-optimal $e$-values and $e$-value-based design, as well as an analysis of the resulting policies. In particular, we discretize the bets and~$e$-space for the Markov chain in~\eqref{eq:transition}, which allows us to use solution techniques for discrete Markov decision processes~\citep{puterman1994markov}.

\subsection{Blocked designs}\label{remark:n_interims}
Due to logistical restrictions~(e.g., fast participant recruitment or a long time to observing responses), it is sometimes not possible to conduct (or at least to plan for) a clinical trial where participant outcomes are analyzed when they come in, leading to a design where participant outcomes are analyzed in blocks, i.e., a blocked design. In fact, 
\citet{Grayling2022estimation} state that the most common single-arm phase II design is Simon's two-stage design,
which has one interim and one final analysis. 

In a blocked design, a fully sequential betting strategy can still be employed for blocks of participants when adhering to the order in which the patients observe their outcomes. In the betting analogy, we can see this as the bettor waiting with their play until the next block of outcomes can be analyzed~(at the interim analysis point). 
In this setting,~$H_0$ is rejected whenever the maximum capital calculated in a fully sequential manner exceeds a threshold~$1/\alpha$, which should still yield type I error rate control due to~\eqref{VRineq}. In addition, the trial is stopped for futility in case the minimum $e$-value calculated in a fully sequential manner reaches the hopeless zone (including zero), or a futility stopping decision is made. 
When analyzing outcomes fully sequentially in blocks, the power stays the same as in a design with a fully sequential, non-blocked design. Hence, the P-max betting strategy remains the same under a fully sequential and blocked design.
In particular, as the power is invariant under the block configuration, we conclude that this analysis approach guarantees a minimum power, no matter if the originally planned interim analysis points are adhered to, so long as the maximum trial size remains the same. 
When analyzing outcomes fully sequentially in blocks, stopping is only possible after the outcomes for a block of participants have been collected. Hence, letting~$n_t,\bar{n}_t$ be the block size for participant~$t$ and sum of block sizes up to and including the block for participant~$t$, in the blocked setting the rewards~\eqref{eqn:rewards_ESS_min_e} have to be adjusted to~$r(\bX_t, B_{t+1})=-n_{t+1}\mathbb{I}(m(\bX_t)<1/\alpha, \bar{n}_t=t)$ to indicate that the next block of participants is recruited whenever the null hypothesis is not rejected at an interim analysis.  

We note that the ``fully sequential in blocks'' approach sketched above is not the only approach for constructing $e$-values on blocked data. One approach would be to treat the blocked data as a binomial outcome and to keep the bet fixed in blocks; however, we found that this approach leads to substantial decreases in power over a fully sequential approach when the block sizes become large. Another approach is to use the ``sequential in blocks'' approach, but to take the average over $e$-values computed from permutations of the outcome sequence, which is more difficult to analyze than the ``fully sequential in blocks'' approach, but could warrant future research.

\subsection{Why use $e$-values in adaptive designs?}\label{sect:reasons_e_design}
This section comments on why $e$-values could be preferred over the current state-of-the-art in adaptive clinical trial designs.

\begin{itemize}
    \item {\bf Not necessary to specify monitoring schedule upfront}\\
    Current single-arm designs assume a pre-specified monitoring schedule that, when adhered to, results in exact type I error rate control. In practice, it may be difficult to exactly adhere to this monitoring schedule (e.g., due to unexpected dropout, delays in collecting responses, or difficulties with recruitment). While methods exist to circumvent issues with type I error rate control in such settings~\citep[see, e.g.,][ for the case of Simon's two-stage design]{koyamaproper2008}, $e$-values are naturally designed to preserve type I error rate control under irregular and non-prespecified interim looks at the data. For instance, this property makes $e$-values also allow for time-based interim analyses. We note that, although this does not have to be the case for power,  any unforeseen change in design may lead to a loss of global optimality of the design-based $e$-values defined in this section. However, the design-optimal $e$-value can be ``re-optimized'', by re-solving~\eqref{opt_prob} from the interim until the end of the trial, given the new design configuration.
    \item {\bf Combining evidence and optional continuation of the trial}\\
    As stated in Section~\ref{sect:e_vals}, the multiplication of two independent $e$-values yields another~$e$-value. This makes it possible to (i) start the trial with an $e$-value from a previous trial, using historical data effectively; (ii) continue the trial and collect a few more observations, which may even be based on the $e$-value observed after all data has been collected~(as long as the $e$-process remains a test martingale, the inference is valid). In addition, while not considered here, in settings where the alternative hypothesis is composite,  historical data may be incorporated in the $e$-value without loss of type I error rate control by combining prior data in ``plug-in'' approaches~\citep{grunwald2024}. 
    \item {\bf Quantifying evidence against the null}\\
    $e$-values inherently quantify evidence against the null hypothesis, and can be transformed to (anytime-valid) p-values by taking their reciprocal. Adaptive designs based on $e$-values hence have the advantage that a measure of evidence against the null hypothesis is readily available, something which is not guaranteed in adaptive designs~(e.g.,~\citet{koyamaproper2008} mention two possible $p$-values based on Simon's two-stage design).
    This also means that, similar to conditional power and posterior probability of a trial success, an~$e$-value indicates how likely it is that the efficacy boundary will be reached during the trial. The benefit of using $e$-values is that conditional power and posterior success probabilities require additional computation, while the~$e$-value can be used for both efficacy declarations and indicating futility of a trial's continuation.
    In addition, as they quantify evidence, $e$-processes can often be transformed to anytime-valid confidence sequences by inversion~\citep{grunwald2024}.   
    \item {\bf Non-binding futility stopping}\\
    As indicated above, we can integrate adaptive clinical trial design with $e$-values by basing futility stopping decisions on $e$-values. The futility stopping decisions do not necessarily have to be binding, meaning that in an adaptive design based on $e$-values, the futility stopping decisions need not be adhered to, without loss of type I and II error rate control. This is not the case for Simon's two-stage design, or designs based on stochastic curtailment~(as in \citet{law2019optimalcurtaileddesignssingle}). We do note that in our $e$-value-based designs, a strictly binding curtailment decision is reached when the capital runs out, which is a scenario that can be mitigated by not allowing~$B_t=1$. However, we note that the situation in which the capital runs out might also be preferable to a situation where the capital is (unbeknownst to the investigators), infinitesimally small and hence ``effectively'' equal to zero.
\end{itemize}

\FloatBarrier
\section{Numerical comparison: Standard and~$e$-value-based multi-stage single-arm designs}\label{sect:OC_results}
\FloatBarrier

In this section, we compare operating characteristics of the $e$-values introduced in Section~\ref{sect:design_optimal} with those of a single-arm stochastic curtailment (SC) design stopping for futility and efficacy~\citep{law2019optimalcurtaileddesignssingle}.
The SC design, found through the \verb|R| package \verb|curtailment|, stops early when the conditional power of a one-sided binomial test~exceeds or goes below pre-specified levels.  Optimal designs using this stopping rule can be found for any number of interim analyses. To make the comparison to $e$-value-based designs fair, we keep the interim analysis points and maximum sample size fixed for SC. The objective for the SC design was to minimize the expected sample size under~$\theta=\theta_1$, and the minimum power for the SC design and $e$-value-based design was set to~$0.8$. 

The operating characteristics we investigate are the cumulative type I error rate and power, and cumulative futility stopping probability under~$\theta\in\{\theta_0,\theta_1\}$. The cumulative type I error rates and power at time~$t$ are the probability of a false or true rejection, respectively, of the null hypothesis before or after observing outcome~$t.$ The cumulative futility stopping probability at time~$t$ is the probability of futility stopping before or at time~$t$, where for the design-optimal $e$-values and $e$-value-based design, we assume that the trial is stopped for futility when a futility stopping decision is made in the $e$-value-based design or when the~$e$-value is in the HZ~(see Remark~\ref{remark:hopeless_almost}).
The operating characteristics are computed using forward recursion on the Markov chain in~\eqref{eq:disc_transition_M}, approximating the Markov chain in~\eqref{eq:transition}. We inspect designs with analyses after every participant, as well as blocked designs. For the blocked designs, we assume a fully sequential analysis is performed at the interim analysis times~(see Section~\ref{remark:n_interims}).

\begin{figure}[htb!]
    \centering
    \includegraphics[width=.95\linewidth]{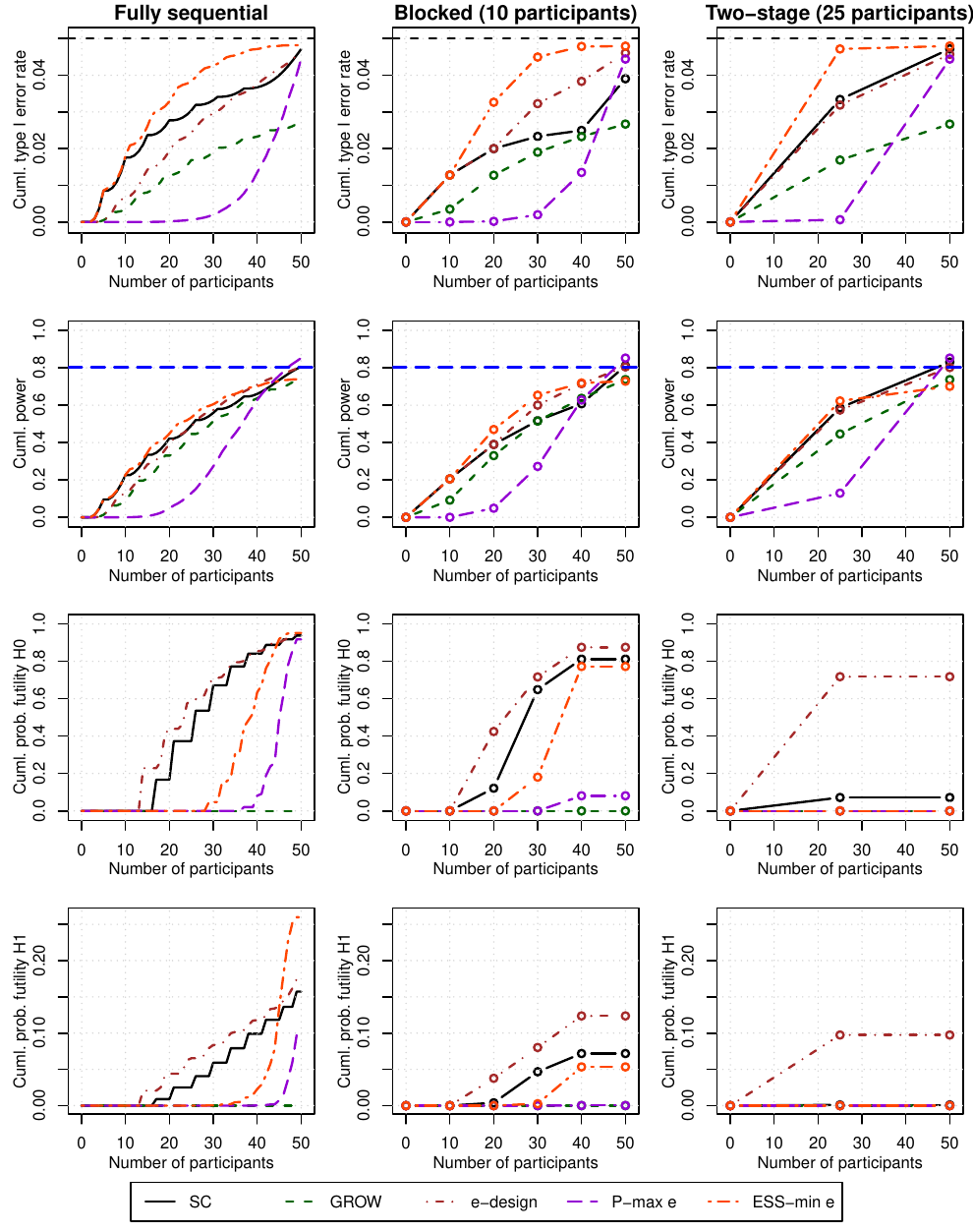}
    \caption{Cumulative type I error rate~(first row), power~(second row), probability of futility stopping when~$\theta=\theta_0$ (third row), and probability of futility stopping when~$\theta=\theta_1$ (fourth row) for the stochastic curtailment~(SC) design, GROW $e$-value,  $e$-value-based design~($e$-design), the power-maximizing~(P-max) $e$-value, and expected sample size minimizing~(ESS-min) $e$-value for a fully sequential trial~(left, analysis after every participant), a blocked design with interim analyses after every ten participants~(middle column), and two-stage design with equal block sizes~(right). The analysis points are shown using dots. The dash-dot-long-dashed line in the second row denotes the power under the non-sequential binomial test for~$\theta=\tilde{\theta}_1=0.242$ versus~$H_0:\theta\leq\theta_0=0.1$. The minimum power was set to 0.8 for the SC and $e$-value-based designs.}
    \label{fig:comparison_OCs}
\end{figure}
\FloatBarrier
For our comparison, we choose a setting where the maximum trial size $n$ equals~$50$ participants, the maximum success rate $\theta_0$ under~$H_0$ is~$0.1$, while the lowest success rate~$\theta_1$ under~$H_1$ is~$0.242$ (chosen to give a power of~$\sim80$\% for the binomial test), lastly, the nominal significance level~$\alpha$ is set to~$0.05$.
Low success rates~$\theta_0$ under~$H_0$, between~$0.05$ and~$0.3$, are common for applications in oncology where the outcome is tumor response, see, e.g.,~\citet{MANDER2010572}. Returning to the betting analogy, the capital of the gambler can at most be multiplied by~$1/\theta_0=10$~(in case $B_{t}=1$), hence, it is no longer possible to reach the threshold~$1/\alpha=20$ for~$\hat{M}_{t}< 20\cdot 10^{t-50}.$ 

Figure~\ref{fig:comparison_OCs} shows the cumulative design operating characteristics for a fully sequential, blocked, and two-stage design.
The efficacy results are consistent over stage sizes.
As expected, the cumulative type I error rates remain below the nominal significance level, where the SC and non-GROW $e$-value-based designs reach a final type I error rate close to the nominal level. This happens relatively early on for the ESS-min betting strategy.
In terms of cumulative power, the GROW $e$-value is dominated by SC, the ESS-min $e$-value, and the $e$-value-based design. Instead, the P-max $e$-value-based design has a low early power, while ending up at a higher power value than SC. This downside of the P-max $e$-value was also noted in~\citet{clerico2026time}.
The ESS-min $e$-value-based design has a lower median expected sample size under~$\theta=\theta_1$ than SC. Across batch sizes, the $e$-value-based design has a similar ESS and final power as SC.  
  To conclude, while implementing the (standard) GROW $e$-value leads to worse operating characteristics than standard adaptive designs~(SC in our case),  design-optimal $e$-values can lead to comparable or even better operating characteristics, while furthermore ensuring anytime validity. In addition, we note that, across batch sizes, the SC, P-max, and $e$-value-based designs reach a power equal to or above that of the one-analysis case (binomial test) while having a lower expected sample size under~$H_1$. 

The two bottom rows of Figure~\ref{fig:comparison_OCs} show that the futility stopping probability curves look quite different across block sizes.
Note that the cumulative futility stopping probability stays constant for the last analysis point, as no early stopping can occur there (nevertheless, we want to show the same horizontal range across figures). For a fully sequential design, the futility stopping curves for SC and the $e$-value-based design look similar, with the $e$-value-based design showing a better performance, but move further apart when increasing the block size. For the two-stage design, the probability of stopping under~$H_0$ and~$H_1$ is only large (e.g., around 70\%) for the $e$-value-based design, whereas the probability is below 10\% for the other designs. Curiously, the probability that the ESS-min $e$-value-based design stops for futility under~$H_1$ becomes relatively large towards the end, which might be a result of the betting strategy becoming more aggressive towards the end when the probability of ever reaching the efficacy boundary is small. By default, futility stopping does not occur for the GROW $e$-value/Kelly betting. In conclusion, we see that the $e$-value-based design yields sufficient power and shows a higher probability of early termination under the null than SC, even for a small number of interim analyses.  

\FloatBarrier

\section{Discussion}\label{s:discuss}
We have investigated multi-stage single-arm clinical trials with binary data based on~$e$-values.
In comparison to more conventional adaptive designs, such designs provide more robustness to type I error rate inflation under trial protocol deviations and trial continuation.

As an alternative to stopping rules based on Kelly betting/GROW $e$-processes, which are asymptotically optimal, we consider betting strategies that maximize the objective~(e.g., power) at a finite horizon. First, we consider betting procedures that either maximize the power in the trial or minimize the expected trial size.  
Second, we consider the idea of $e$-value-based  adaptive designs, and illustrate this idea by constructing single-arm adaptive designs with a minimum power constraint and optimal futility stopping, where the efficacy boundary is based on the~$e$-value exceeding the threshold~$1/\alpha$.
We construct these $e$-value-based designs by dynamic programming, using a discretization of the~$e$-space of the  Markov decision process defined by the $e$-value-based design. We propose to analyze data fully sequentially in blocks, a procedure that ensures the minimum power is attained, even if the realized block sizes differ from the originally planned block sizes, so long as the maximum trial size remains the same.

The main takeaway from our research is that, by not adhering to the  GROW $e$-process default in small samples, $e$-value-based designs can be made highly competitive with standard single-arm adaptive designs with binary data, even when reducing the number of analyses from a fully sequential design to a two-stage design. 
Our numerical results indicate that, when maximizing power, $e$-value-based designs can lead to higher power than adaptive and standard (fixed sample) binomial tests, while having superior flexibility in terms of non-adherence to trial protocols. When minimizing expected sample size, we recommend using the $e$-value-based design, which enforces the trial to have sufficient power while resulting in a high probability of futility stopping under the null hypothesis. 
One additional major benefit of design-optimal $e$-values is that they more readily indicate situations where non-stochastic curtailment should be applied: in cases where the $e$-value is zero or reaches the ``hopeless zone'' (Section~\ref{sect:comp_policy_analysis}), the trial can be stopped for futility. In case the $e$-value reaches zero, the futility stopping decision is, by definition, binding as, not entirely dissimilar to a setting with non-stochastic curtailment, the conditional power is zero. We note that, if this feature is undesired, the bets can be restricted to the interval~$[0,1)$ and in this setting the $e$-value can never reach the value zero. However, starting a completely new experiment with starting $e$-value 1 can be more efficient than optional continuation with very small, but nonzero, $e$-values.

A downside of design-based $e$-values is that they lose the connection to likelihood ratios that many GROW $e$-values have. While any design-optimal $e$-value larger than 1 can be interpreted as evidence against the null hypothesis, $e$-values smaller than 1 cannot be directly interpreted as evidence against the alternative. Rather, small design-optimal $e$-values indicate futility of continuing the experiment; curtailment by a zero $e$-value expresses that your maximum sample-size experiment is pointless.

 In the current paper, we focused on testing instead of estimation. We note that testing obtains a higher emphasis than estimation in multi-stage single-arm designs, as these designs are often conducted in phase II, with a larger trial being conducted in phase III, where estimation obtains higher emphasis. Nevertheless, estimation may still be of relevance for these designs. 
 Considering point estimation, the standard maximum likelihood estimator will be biased, and any available bias-adjusted estimator may still be used in $e$-value-based designs, with methods listed in Section~2.9 of~\citet{law2019optimalcurtaileddesignssingle}. Another field of interest for future research is confidence interval construction, in particular, how to construct a design-optimal anytime-valid confidence interval for the success rate in single-arm designs. We note that, for trials with a small number of interims, continuous monitoring of the estimator along with its uncertainty (which is the main appeal of this method) is not realistic, but having an anytime-valid confidence interval may still be useful due to robustness against design violations.

Several areas are left for future research.  Simon's two-stage design typically minimizes the expected trial size under the null hypothesis while controlling the trial's power. This mixture of hypotheses within the optimization problem is currently not covered, but one way to solve such optimization problems would be the approach in~\citet{baas2024CMDP}, where constrained Markov decision processes are considered with constraints under different statistical hypotheses. 
In addition to this, future research could also focus on minimizing the maximum trial size and optimizing the interim analysis points.
The current paper focuses on single-arm designs as a starting point for $e$-value-based designs, but the idea of $e$-value-based adaptive designs can be applied in more general settings. Immediate directions of interest are trials with multiple treatments and trials using response-adaptive randomization~\citep{robertson2023response}, which often suffer from type I error rate inflation. This would lead to a more natural method of constructing exact group sequential designs for response-adaptive designs than the approach in~\citet{BAAS2025108207}, which mainly focused on the one-stage setting, while the $e$-value approach also allows for time-based interim analyses. 
In such settings, two-sided testing could also be investigated, which remains an open problem. 
For randomized clinical trials, the combination of $e$-value-based inference and randomization-based inference~\citep{rosenberger2019randomization} would also be of interest, to construct efficient sequential methods of statistical inference robust to both protocol deviations and unmeasured confounders.
Unlike p-values, $e$-values do not require the data distribution to be fully known under the null hypothesis, as~$e$-values are defined through their expectation under the null. This property is especially advantageous in complex adaptive designs where ad-hoc design changes can occur, and constructing a representative data distribution is difficult. For example, this property could be of use in master protocols, where treatments are added whenever they are discovered. \citet{robertson2023online} consider the application of approaches with anytime validity in such trials, and we note that the GROW $e$-process may be preferred over our design-optimal approach in such perpetual trials.  
As the state space of the Markov decision process in Section~\ref{sect:comp_policy_analysis} does not change when considering other outcomes from a one-parameter exponential family~(e.g., Poisson outcomes or normal outcomes with known variance), our approach is readily generalizable to designs with other outcome types than binary, which would also be an interesting direction for future research.
As discussed in Section~\ref{sect:reasons_e_design}, $e$-value-based designs can allow for the incorporation of historical data without type I error rate inflation. Such designs can provide a major benefit over traditional Bayesian designs with historical borrowing that often suffer from type I error rate inflation, which is a hurdle for regulators~\citep{lim2018minimizing}. Hence, an interesting future research direction could consider $e$-value-based designs using historical borrowing.

\bibliographystyle{unsrtnat}
\bibliography{main.bib}     %

@article{koyamaproper2008,
author = {Koyama, Tatsuki and Chen, Heidi},
title = {Proper inference from {S}imon's two-stage designs},
journal = {Statistics in Medicine},
volume = {27},
number = {16},
pages = {3145-3154},
url = {https://doi.org/10.1002/sim.3123},
year = {2008}
}

@incollection{breimualphanu2011optimal,
    AUTHOR = {Breiman, Leo},
     TITLE = {Optimal gambling systems for favorable games},
 BOOKTITLE = {Proc. 4th {B}erkeley {S}ympos. {M}ath. {S}tatist. and {P}rob.,
              {V}ol. {I}},
     PAGES = {65--78},
 PUBLISHER = {Univ. California Press, Berkeley-Los Angeles, Calif.},
      YEAR = {1961},
   MRCLASS = {90.72},
  MRNUMBER = {135630},
MRREVIEWER = {E.\ S.\ Keeping},
}

@article{pallmann2018adaptive,
  title={Adaptive designs in clinical trials: {W}hy use them, and how to run and report them},
  author={Pallmann, Philip and Bedding, Alun W. and Choodari-Oskooei, Babak and Dimairo, Munyaradzi and Flight, Laura and Hampson, Lisa V. and Holmes, Jane and Mander, Adrian P. and Odondi, Lang’o and Sydes, Matthew R. and Villar, Sof\'ia S. and Wason, James M. S. and Weir, Christopher J. and Wheeler, Graham M. and Yap, Christina  and Jaki,  Thomas},
  journal={BMC Medicine},
  volume={16},
  number={29},
  pages={1-15},
  year={2018},
  url={https://doi.org/10.1186/s12916-018-1017-7}
}

@article{koning2026anytime,
    author = {Koning, Nick W. and van Meer, Sam},
    title = {Anytime validity is free: {I}nducing sequential tests},
    journal = {Journal of the Royal Statistical Society Series B: Statistical Methodology},
    pages = {1-19},
    year = {2026},
    url = {https://doi.org/10.1093/jrsssb/qkag050}
}

@phdthesis{ville1939etude,
    author = {Jean Ville},
    title = {Etude critique de la notion de collectif},
    school = {Gauthier-Villars, Paris},
    year = {1939},
    url = {https://www.numdam.org/issue/THESE_1939__218__1_0.pdf}
}

@article{chi2008curtailed,
  title={Curtailed two-stage designs in phase {II} clinical trials},
  author={Chi, Yunchan and Chen, Chia-Min},
  journal={Statistics in Medicine},
  volume={27},
  number={29},
  pages={6175-6189},
  year={2008},
  url = {https://doi.org/10.1002/sim.3424}
}

@article{ayanlowo2007stochastically,
  title={Stochastically curtailed phase {II} clinical trials},
  author={Ayanlowo, Ayanbola O. and Redden, David T.},
  journal={Statistics in medicine},
  volume={26},
  number={7},
  pages={1462-1472},
  year={2007},
  url={https://doi.org/10.1002/sim.2653}
}

@book{altman1999constrained,
  title={{Constrained Markov decision processes}},
  author={Altman, Eitan},
  year={1999},
edition = {first},
  publisher={Routledge, New York}
}

@article{robertson2023online,
  title={Online multiple hypothesis testing},
  author={Robertson, David S. and Wason, James M. S. and Ramdas, Aaditya},
  journal={Statistical science},
  volume={38},
  number={4},
  pages={557},
  year={2023},
  url = {https://doi.org/10.1214/23-STS901}
}

@article{lim2018minimizing,
  title   = {Minimizing Patient Burden Through the Use of Historical Subject-Level Data in Innovative Confirmatory Clinical Trials: {R}eview of Methods and Opportunities},
  author  = {Lim, Jenny and Walley, Robert and Yuan, Ji and Liu, C. and Wright, D. and Frattini  et al., M.},
  journal = {Therapeutic Innovation \& Regulatory Science},
  volume  = {52},
  number  = {5},
  pages   = {546-559},
  year    = {2018},
 url     = {https://doi.org/10.1177/2168479018778282}
}

@article{robertson2023response,
	title={{Response-adaptive randomization in clinical trials: From myths to practical considerations}},
	author={Robertson, David S. and Lee, Kim May and L{\'o}pez-Kolkovska, Boryana C. and Villar, Sof{\'\i}a S.},
	journal={Statistical Science},
	volume={38},
	number={2},
	pages={185-208},
	year={2023},
	url = {https://doi.org/10.1214/22-STS865}
}

@article{baas2024CMDP,
      title={{Constrained Markov decision processes for response-adaptive procedures in clinical trials with binary outcomes}}, 
      author={Stef Baas and Aleida Braaksma and Richard J. Boucherie},
      year={2025},
volume = {\hspace{-3.8mm}\phantom{x}},
      journal={Annals of Operations Research},
pages = {1-51},
      url={https://doi.org/10.1007/s10479-025-06703-8}, 
}

@article{shafer2021testingbetting,
    author = {Shafer, Glenn},
    title = {Testing by Betting: {A} Strategy for Statistical and Scientific Communication},
    journal = {Journal of the Royal Statistical Society Series A: Statistics in Society},
    volume = {184},
    number = {2},
    pages = {407-431},
    year = {2021},
    url = {https://doi.org/10.1111/rssa.12647},
}

@article{eltriki2024adaptive,
author = {Ben‑Eltriki, Mohamed  and Rafiq, Aisha and
Paul, Arun and
Prabhu, Devashree and
Afolabi, Michael O. S. and
Baslhaw, Robert and Neilson, Christine J. and
Driedger, Michelle and
Mahmud, Salaheddin M. and
Lacaze‑Masmonteil, Thierry and
Marlin, Susan and
Offringa, Martin and Butcher, Nancy and Heath, Anna  and
Kelly, Lauren E. 
},
title = {Adaptive designs in clinical trials:{A}  systematic review-part {I}},
journal = {BMC Medical Research Methodology},
volume = {24},
number = {229},
year = {2024},
pages = {1-14},
url = {https://doi.org/10.1186/s12874-024-02272-9}
}

@article{taga2026learning,
  title={Learning to Bet for Horizon-Aware Anytime-Valid Testing},
  author={Taga, Ege Onur and Oymak, Samet and Shekhar, Shubhanshu},
  journal={arXiv preprint arXiv:2603.19551},
  year={2026},
  url = {https://arxiv.org/abs/2603.19551}
}

@article{voravcek2025star,
  title={{ST}a{R}-Bets: {S}equential Target-Recalculating Bets for Tighter Confidence Intervals},
  author={Vor{\'a}{\v{c}}ek, V{\'a}clav and Orabona, Francesco},
  journal={arXiv preprint arXiv:2505.22422},
  year={2025},
  url= {https://arxiv.org/abs/2505.22422}
}

@article{shafer2011testmartingales,
author = {Glenn Shafer and Alexander Shen and Nikolai Vereshchagin and Vladimir Vovk},
title = {Test Martingales, {B}ayes Factors and p-Values},
volume = {26},
journal = {Statistical Science},
number = {1},
publisher = {Institute of Mathematical Statistics},
pages = {84-101},
year = {2011},
URL = {https://doi.org/10.1214/10-STS347}
}

@article{ramdas2023game,
  title={Game-theoretic statistics and safe anytime-valid inference},
  author={Ramdas, Aaditya and Gr{\"u}nwald, Peter and Vovk, Vladimir and Shafer, Glenn},
  journal={Statistical Science},
  volume={38},
  number={4},
  pages={576-601},
  year={2023},
 url={https://doi.org/10.1214/23-STS894}
}

@article{law2019optimalcurtaileddesignssingle,
author = {Martin Law and Michael J. Grayling and Adrian P. Mander},
title = {A stochastically curtailed single‐arm phase {II} trial design for binary outcomes},
journal = {Journal of Biopharmaceutical Statistics},
volume = {32},
number = {5},
pages = {671-691},
year = {2022},
URL = { 
    
        https://doi.org/10.1080/10543406.2021.2009498
    
    

}
}

@article{grunwald2024,
    author = {Grünwald, Peter and de Heide, Rianne and Koolen, Wouter},
    title = {Safe testing},
    journal = {Journal of the Royal Statistical Society Series B: Statistical Methodology},
    volume = {86},
    number = {5},
    pages = {1091-1128},
    year = {2024},
    url = {https://doi.org/10.1093/jrsssb/qkae011}
}

@article{WaudbySmithbettingmeans,
    author = {Waudby-Smith, Ian and Ramdas, Aaditya},
    title = {Estimating means of bounded random variables by betting},
    journal = {Journal of the Royal Statistical Society Series B: Statistical Methodology},
    volume = {86},
    number = {1},
    pages = {1-27},
    year = {2024},
    url = {https://doi.org/10.1093/jrsssb/qkad009}
}

@book{puterman1994markov,
	title={{Markov decision pocesses: Discrete stochastic dynamic programming}},
	author={Puterman, Martin L.},
edition = {first},
year = {1994},
url = {https://doi.org/10.1002/9780470316887},
publisher = {Hoboken, NY: John Wiley \& Sons}
	}

@article{johnson2013uniformly,
  title={Uniformly most powerful {B}ayesian tests},
  author={Johnson, Valen E.},
  journal={Annals of statistics},
  volume={41},
  number={4},
  pages={1716-1741},
  year={2013},
  url = {https://doi.org/10.1214/13-AOS1123}
}

@book{berry2010bayesian,
  title={Bayesian Adaptive Methods for Clinical Trials},
  author={Berry, Scott M. and Carlin, Bradley P. and Lee, J. Jack and M\"uller, Peter},
  year={2010},
  publisher={Chapman \& Hall/CRC}
}

@article{BAAS2025108207,
title = {Exact statistical analysis for response-adaptive clinical trials: {A} general and computationally tractable approach},
journal = {Computational Statistics \& Data Analysis},
volume = {211},
pages = {108207},
year = {2025},
url = {https://doi.org/10.1016/j.csda.2025.108207},
author = {Stef Baas and Peter Jacko and Sof\'ia S. Villar}
}

@article{schure2022allin,
   author = {{t}er Schure, Judith and Grünwald, Peter},
title = {{ALL-IN} meta-analysis: {B}reathing life into living systematic reviews and prospective meta-analyses},
journal = {F1000Research},
year = {2025},
volume = {11},
number = {549},
pages = {1-31},
url = {https://doi.org/10.12688/f1000research.74223.2}
}

@book{jennison2000group,
  title={Group Sequential Methods with Applications to Clinical Trials},
  author={Jennison, Christopher and Turnbull, Bruce W.},
  year={2000},
  publisher={Chapman \& Hall/CRC}
}

@article{Grayling2022estimation,
author = {Michael J. Grayling and Adrian P. Mander},
title = {Optimised point estimators for multi-stage single-arm phase {II} oncology trials},
journal = {Journal of Biopharmaceutical Statistics},
volume = {32},
number = {6},
pages = {817-831},
year = {2022},
URL = { 
    
        https://doi.org/10.1080/10543406.2022.2041656
    
    

}
}

@ARTICLE{kelly_new_1956,
  author={Kelly, John L.},
  journal={The Bell System Technical Journal}, 
  title={A new interpretation of information rate}, 
  year={1956},
  volume={35},
  number={4},
  pages={917-926},
  doi={10.1002/j.1538-7305.1956.tb03809.x}}

@article{MANDER2010572,
title = {Two-stage designs optimal under the alternative hypothesis for phase {II} cancer clinical trials},
journal = {Contemporary Clinical Trials},
volume = {31},
number = {6},
pages = {572-578},
year = {2010},
issn = {1551-7144},
url = {https://doi.org/10.1016/j.cct.2010.07.008},
author = {Adrien P. Mander and Simon G. Thompson},
}

@article{wald1945sequential,
  title={Sequential tests of statistical hypotheses},
  author={Wald, Abraham},
  journal={The Annals of Mathematical Statistics},
  pages={117-186},
volume = {16},
number ={2},
  year={1945},
  url = {https://www.jstor.org/stable/2235829}
}

@article{rosenberger2019randomization,
author = {Rosenberger, William F. and Uschner, Diane and Wang, Yanying},
title = {Randomization: {T}he forgotten component of the randomized clinical trial},
journal = {Statistics in Medicine},
volume = {38},
number = {1},
pages = {1-12},
doi = {https://doi.org/10.1002/sim.7901},
year = {2019}
}

@ARTICLE{fisher_ramdas2026,
  author={Fischer, Lasse and Ramdas, Aaditya},
  journal={IEEE Transactions on Information Theory}, 
  title={Improving {W}ald’s (Approximate) Sequential Probability Ratio Test by Avoiding Overshoot}, 
  year={2026},
  volume={72},
  number={4},
  pages={2457-2471},
  url = {https://doi.org/10.1109/TIT.2026.3658855}
 }

@article{ramdas2025hypothesis,
    author = {Ramdas, Aaditya and Wang, Ruodu},
    title = {Hypothesis Testing with E-values},
    journal = {Foundations and Trends in Statistics},
    volume = {1},
    number = {1-2},
    pages = {1-390},
    year = {2025},
    month = {07},                
    url = {https://doi.org/10.1561/3600000002},
}

@article{clerico2026time,
  title={Time-sensitive anytime-valid testing},
  author={Clerico, Eugenio and Wegel, Tobias and Azangulov, Iskander and Rebeschini, Patrick},
  journal={arXiv preprint arXiv:2605.06521},
  year={2026},
  url = {https://arxiv.org/abs/2605.06521}
}
\par\vspace{.5em} %
\FloatBarrier
\appendix

\section{Computational details and betting strategy analysis}\label{sect:comp_policy_analysis}
\subsection{Design-optimal betting strategies}\label{sect:DO_policy_analysis}
   To compute the design-optimal $e$-value, we discretize the range of possible  $e$-values to a finite set~$\hat{\mathcal{M}}\subset[0,1/\alpha]$ which includes~$0$, $1$ and~$1/\alpha$ as its elements. 
   Furthermore, we discretize the bet sizes~$B_t$ to a set~$\hat{\mathcal{B}}\subset[0,1]$.
   Let~$\ubar{\Pi}_{\hat{M}}(x)=\max\{m\in\hat{M}:m\leq x\}$ be the largest element in~$\hat{M}$ smaller than~$x.$
   We let~$\hat{\bX}_t$ denote the state of the discretized Markov decision process (i.e., the discretized version of~$\bX_t$), which has the following updating rule
   \begin{equation}(t,\hat{M}_t)\stackrel{B_{t+1}}{\rightarrow}
    (t+1,\,\ubar{\Pi}_{\hat{M}}(\hat{M}_t\cdot (1+B_{t+1}(Y_{t+1}/\theta_0-1)))).\label{eq:disc_transition_M}\end{equation}
    The $e$-values~$\hat{M}_t\in\{0,1/\alpha\}$ again lead to absorbing states.
    The rewards~$r(\bX_t, B_t)$ for placing bet~$B_t$ in state~$\bX_t$ remain the same, and we solve~\eqref{opt_prob} using the states, actions, and rewards for the discretized problem. For our computations, we used a combination of \verb|R| (version 4.5.3) and \verb|C++| using the \verb|Rcpp| R-package.  
    
    Let~$(\hat{M}_t)_t$ be the discrete $e$-process following from~\eqref{eq:disc_transition_M} and~$(\tilde{M}_t)_t$ be the continuous $e$-process following from~\eqref{capital} where the bets~$B_t$ follow from a betting strategy in the discretized problem stated above with input states~$\hat{M}_t$ based on the same outcomes. By induction, it follows that~$\tilde{M}_t\geq \hat{M}_t$ for all~$n$ and hence $ \hat{M}_t\geq 1/\alpha \implies \tilde{M}_t\geq1/\alpha$ meaning that the power value~(expected sample size) found under~$\hat{M}_t$ is a lower (upper) bound for this value under~$\tilde{M}_t$. Note that type I error rate control is still guaranteed under~$\tilde{M}_t$, as this process remains a test supermartingale under~$H_0.$

\begin{figure}[htb]
    \centering
    \includegraphics[width=\linewidth]{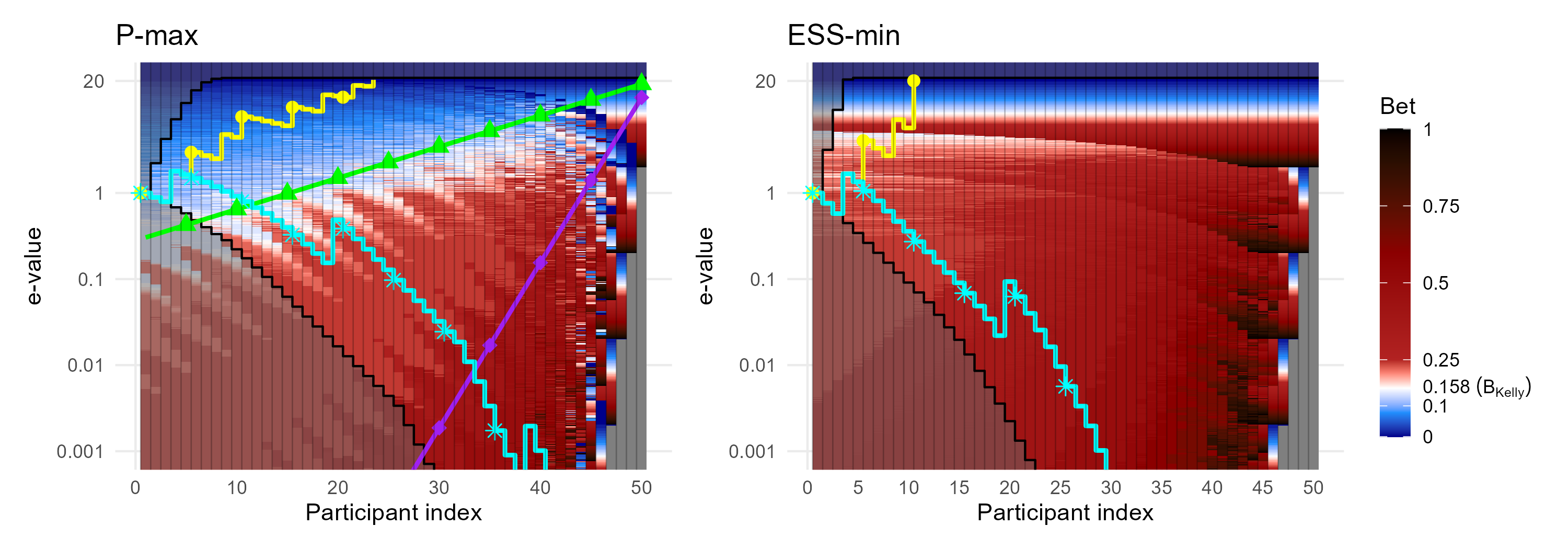}
    
    \caption{$n=50$, $\theta_0=0.1$, $\theta_1=0.242$, $\alpha=0.05$. Left: optimal bets~$B^{\text{P-max}}_t$ for the power-maximizing $e$-value. The~(green) line with triangle markers indicates the $e$-values that exactly match the bound~$1/\alpha$ at the final interim under the growth rate realized under Kelly betting, calculated through~\eqref{eq:ub}, which corresponds to a bet equal to~$B_{\text{Kelly}}\approx 0.158$. The~(purple) line with diamond markers indicates $e$-values under which the GROW $e$-process exactly matches the level~$1/\alpha$ after $(n-t+1)/2$ successes~\citep[see][]{taga2026learning}. Right: optimal bets~$B^{\text{ESS-min}}_t$ for the expected sample-size minimizing $e$-value. For both subfigures, the shaded region indicates $e$-values that are impossible to reach starting at an initial $e$-value~$M_0=1$ under the optimal betting strategies.
    For both subfigures, the hopeless zone~(Remark~\ref{remark:hopeless_almost}) is indicated in gray.
    The (yellow) path with round markers indicates a sample path under the alternative hypothesis ($\theta=\theta_1$) and the (cyan) path with asterisks is an $e$-process path under the null hypothesis~($\theta=\theta_0$). Common outcomes are used across figures for each (yellow) path with dot markers and for each (cyan) path with asterisk markers.}
    \label{fig:policy_plots_design_optimal}
\end{figure}

To calculate the design-optimal betting strategies, we set~$\hat{\mathcal{B}}=\{0,0.0001,0.001,0.01,0.02,\dots, 0.99,0.999,0.9999,1\}$ and~$\hat{\mathcal{M}}$ equal to the set containing the value 0, the logarithmic range with 1000 values from $10^{-5}$ until $1-2\epsilon$ (with $\epsilon$ the machine $\epsilon$, in this case about~$2\cdot 10^{-16}$) and an equidistant range from 1 to $1/\alpha$ of 1000 values. Figure~\ref{fig:policy_plots_design_optimal} shows the calculated P-max and ESS-min design-optimal betting strategies for the same trial setting as in Section~\ref{sect:OC_results}.

To compare the P-max betting strategy with the theoretical results found in~\citet{taga2026learning}, we show the $e$-values~$\bar{m}_t$ exactly matching the bound~$1/\alpha$ at the final interim under the growth rate realized under Kelly betting, i.e.,~\begin{equation}\bar{m}_t=\log(1/\alpha) - \mathbb{E}_{\theta_1}[\log(1+B_{\text{Kelly}}(Y_1/\theta_0-1))]/(n - t) \label{eq:ub}\end{equation} as well as the $e$-values $\ubar{m}_t$  matching the level~$1/\alpha$ after $(n-t)/2$ successes under Kelly betting:
$$\ubar{m}_t=\log(1/\alpha) - \log(\theta_1/\theta_0)/(2(n - t)) .$$
\citet{taga2026learning} show that it is better to bet less aggressively than Kelly betting when~$M_t\geq \bar{m}_t$~(Proposition~3.6) while it is better to bet more aggressively than Kelly betting when~$M_t\leq\ubar{m}_t$ (Proposition~3.4). We note that the comparison in the propositions of~\citet{taga2026learning} considers policies that always place a constant bet, while for our design-optimal betting strategies, the bet size will depend on the state~$\bX_t.$ Nevertheless, we find agreement with the results in~\citet{taga2026learning}:    for~$M_t\geq\bar{m}_t$ we have~$B_{t+1}^{\text{P-max}}\leq B_{\text{Kelly}}$ in many settings and for~$M_t\leq \ubar{m}_t$ we often have~$B_{t+1}^{\text{P-max}}\geq B_{\text{Kelly}}$ (here, $B_{t+1}^{\text{P-max}}$ is the P-max bet for~$M_t$). We note that the lower bound~$(\bar{m}_t)_t$ seems to be conservative, in the sense that for many $e$-values~$M_t$ in-between~$\bar{m}_t$ and~$\ubar{m}_t$ we have~$B_{t+1}^\text{P-max}\geq B_{\text{Kelly}}.$

The shaded area in the subfigures of Figure~\ref{fig:policy_plots_design_optimal} corresponds to $e$-values that cannot be reached under the optimal betting strategies for a starting $e$-value~$M_0$ equal to~$1$ (note that this area depends on the betting strategy). 
We see that~$B^{\text{P-max}}_{t}\leq B_{\text{Kelly}}$ is guaranteed for~$k\leq6$, Kelly betting is often optimal around the line~$M_t=\bar{m}_t$, and below this line it is often optimal to have~$B_t> B_{\text{Kelly}}.$ The betting behavior is much more aggressive under the ESS-min strategy: we have~$B_1^{\text{ESS-min}}\geq B_{\text{Kelly}}$ and see that~$B_t^{\text{ESS-min}}\geq B_{\text{Kelly}}$ for many more $e$-values than under the P-max strategy. As a result, the shaded region is smaller for the ESS-min strategy: the gambler can win and go bust earlier. The ESS-min strategy seems to have a more time-homogeneous behavior than the P-max strategy, which makes sense as the costs are roughly divided equally over decision epochs. 

Three regions stand out in Figure~\ref{fig:policy_plots_design_optimal}. First, the HZ~(Remark~\ref{remark:hopeless_almost}) is indicated in gray, as there is no optimal bet size. Second, to the left of the HZ, we have the AHZ, where it is optimal to bet more conservatively to increase the probability of having a relatively high $e$-value when ending up in the HZ.
Third, at the top-right of the betting strategy plots~(top left as well for the ESS-min betting strategy), we see an area where the optimal bet is time-constant. This is the area where the optimal bet equals the one-stage uniformly most powerful~(OS-UMP) bet; we know this from the fact that this bet is optimal at the final interim~(see the discussion around~\eqref{match_rate}). 

Figure~\ref{fig:policy_plots_design_optimal} shows two sample paths for each betting strategy, one under $\theta=\theta_0$~(yellow, dot markers) and one under~$\theta=\theta_1$~(cyan, asterisk markers). The path under~$H_1$ reaches the level~$1/\alpha$ for both plots, and quicker under the ESS-min $e$-value, as might be expected. The $e$-process path under~$H_0$ reaches a small value in both subfigures.

\FloatBarrier
\subsection{$e$-value-based design}
Again, we discretize the range of possible $e$-values, as well as the action space, to compute $e$-value-based designs. To find an optimal constrained betting strategy, using Lemma 9.2 in~\citet{altman1999constrained}, we rewrite~\eqref{ineq:power_constraint} to
$$\max_{\lambda\geq 0} \min_{\pi}\mathbb{E}^{\pi}_{\theta_1}\left[\tau(\pi)\right] + \lambda\left(\mathbb{P}^{\pi}_{\theta_1}\left(m(\bX_t)< 1/\alpha\right)  -\beta\right).$$
For~$\lambda = 0$, the inner minimization step minimizes the expected trial size~(and will hence stop for futility in every state), whereas increasing~$\lambda$ will shift emphasis to finding policies with high power. Policies with an optimal trade-off concentrate at the penalty~$\lambda $ such that~$\mathbb{P}^{\pi}_{\theta_1}\left(m(\bX_t)< 1/\alpha\right)\approx \beta$.
Hence, we apply a discrete Newton method to find~$\lambda$ such that~$\mathbb{P}^{\pi}_{\theta_1}\left(m(\bX_t)< 1/\alpha\right)  <\beta$ and~$|\mathbb{P}^{\pi}_{\theta_1}\left(m(\bX_t)< 1/\alpha\right)  -\beta|\leq \epsilon$~(in our applications, we set~$\epsilon=0.01$).  

\begin{figure}[h!]
    \centering
    \includegraphics[width=0.6\linewidth]{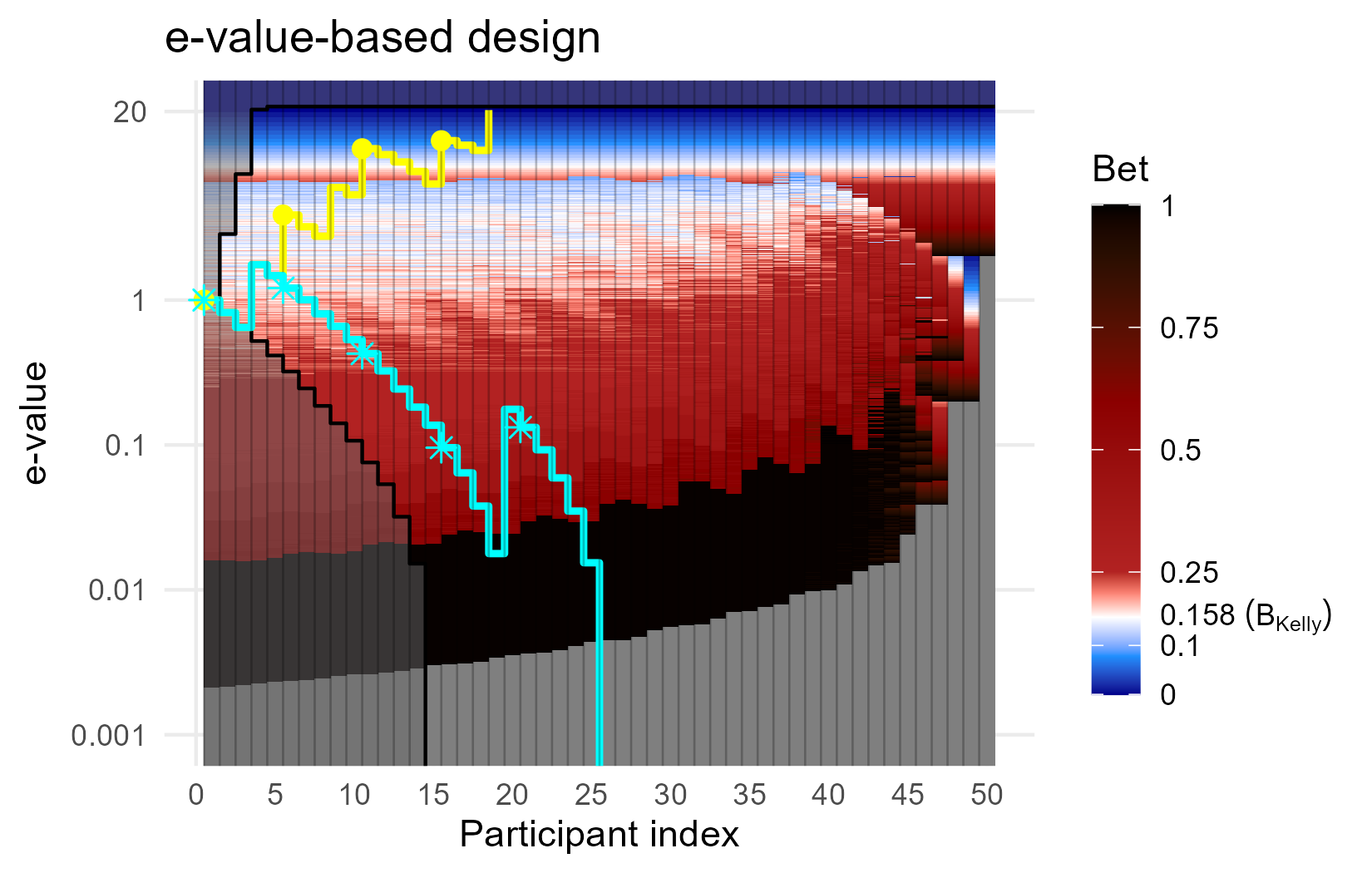}
    \caption{Optimal bets for the $e$-value-based design allowing for stopping for efficacy and futility. The gray cells indicate states where futility stopping is optimal.  The shaded region indicates $e$-values that are impossible to reach starting at an initial $e$-value~$M_0=1$ under the design. $n=50$, $\theta_0=0.1$, $\theta_1=0.242$, $\alpha=0.05$, and $\beta= 0.2$. Same as in Figure~\ref{fig:policy_plots_design_optimal}, the (yellow) path with round markers indicates a sample path under the alternative hypothesis ($\theta=\theta_1$) and the (cyan) path with asterisks is an $e$-process path under the null hypothesis~($\theta=\theta_0$). Both paths use the same outcomes as in Figure~\ref{fig:policy_plots_design_optimal}.}
    \label{fig:policy_plot_fut}
\end{figure}

Figure~\ref{fig:policy_plot_fut} shows the $e$-value-based design found under this procedure~(the sets~$\hat{\mathcal{M}},\hat{\mathcal{B}}$ were chosen the same as in the last section). We see that the initial bet is quite close to the Kelly bet~($B_1=0.161$). In contrast to the ESS-min betting strategy, the betting strategy seems less time-homogeneous, and there are more settings in which it is optimal to bet more conservatively than Kelly betting. In comparison to the P-max $e$-value, the set where it is optimal to use the OS-UMP bet is more similar to the one for the ESS-min $e$-value.  The reason for this is that, given a Lagrange multiplier~$\lambda$, the objective for the optimal design considers a linear combination of power and the ESS, hence the resulting betting strategy will be a mix of the P-max and ESS-min betting strategies. The futility stopping region is monotone and indicated in gray, and as reaching the futility boundary yields a reward in terms of decreasing the sample size, we see that the strategy is betting more aggressively close to the futility stopping boundary, and only for $e$-values close to one it will bet conservatively in the AHZ. Due to this, the lower bound for the capital values also decreases more rapidly than for the ESS-min strategy. 

Looking at the $e$-process paths, the path under~$H_1$ shows a behavior in between those for P-max and ESS-min in Figure~\ref{fig:policy_plots_design_optimal}, while the path under~$H_0$ decreases more quickly than under both design-optimal policies.

\subsection{Distributions of $e$-values and bets over time}
\begin{figure}[h!]
        \centering
        \includegraphics[width=1\linewidth]{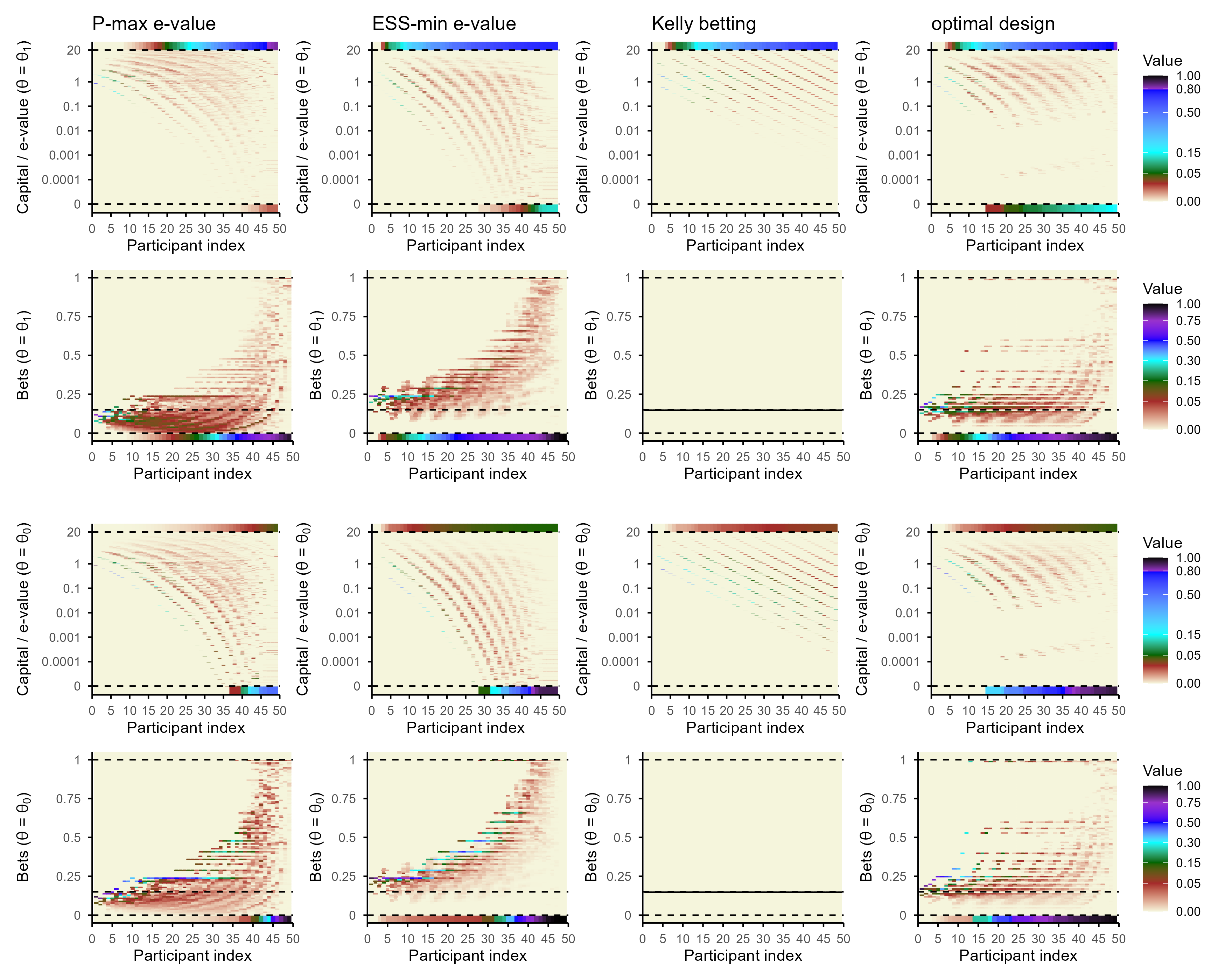}
        \caption{Distribution over time of the  $e$-processes and bets for the P-max, ESS-min, and GROW betting strategies, as well as the $e$-value-based design.  $n=50$, $\theta_0=0.1$, $\theta_1=0.242$, $\alpha=0.05$, and $\beta= 0.2$.   }
        \label{fig:policy_plot_DO}
    \end{figure}

Figure~\ref{fig:policy_plot_DO}  shows the distribution over time of the  $e$-processes and bets for the P-max, ESS-min, and GROW betting strategies, as well as the $e$-value-based design. 
As expected, the P-max $e$-value shows the highest power, above 80\%, at the end of the trial, while the ESS-min $e$-value, GROW $e$-value, and optimal design show a higher power early on. As the bet sizes are fixed, the distribution of the GROW $e$-value is very structured. The optimal design yields a large probability, around~$15\%$ of zero capital around analysis 30. The third row shows the distribution over $e$-values under~$\theta=\theta_0$, which has shifted downwards, as expected, with a type I error rate at most $5\%$ and higher futility stopping probabilities than under~$H_1.$
Looking at the distribution over bet sizes, the bets for all $ e$-value-based designs are seen to become more aggressive towards the end whenever the bets are nonzero, as zero is the bet-size when the trial has stopped for efficacy or futility, with the ESS-min $e$-value having the highest probability mass around high bet sizes. The bets for the optimal e-value-based design have the highest probability of reaching zero before the end of the trial.

\FloatBarrier

\begin{table}[h!]
\section{Terminology table}\label{appendix:terminology_table}
\centering
	{\footnotesize
	\setlength{\tabcolsep}{2mm}	 \renewcommand{\arraystretch}{1.2}\begin{tabular}{m{3.7cm} m{90mm}>{\raggedleft\arraybackslash}p{2.5cm}}\hline
			Term  & Description & Defined in section \\
			\midrule
         Curtailment& Premature stopping of the trial for futility or efficacy when it is (relatively) certain that a futility or efficacy conclusion (resp.) will be reached before the maximum sample size &\ref{sect:introduction} + \ref{sect:analogy}\\
       $e$-value  & Degree of {$\bm e$}vidence against~$H_0$ (capital in betting analogy)  & \ref{sect:e_vals}\\ 
        $e$-process  & Sequence of $e$-values after each data analysis in the trial (capital process in betting analogy)  & \ref{sect:e_vals}\\ 
Anytime-validity& Property under the null hypothesis: Probability that an $e$-value exceeds $1/\alpha>0$ is always less than~$\alpha$. (Note that an $e$-value always equal to zero is useless, but still anytime-valid with zero probability of exceeding any threshold.)&\ref{sect:e_vals}\\
 Bettor & Skeptic of~$H_0$, betting on treatment successes in the trial/statistician designing the $e$-value test  & \ref{sect:analogy}\\
     Kelly betting & Betting strategy to maximize the expected growth rate of the $e$-value (capital) after each analysis &\ref{sect:GROW}\\
          GROW $e$-value/$e$-process & Capital/capital process under Kelly betting, asymptotically optimal &\ref{sect:GROW}\\
             Design-optimal $e$-value & Capital of optimal betting procedure taking the design of the study into account &\ref{sect:DO_e}\\
             $e$-space& Range of possible times and~$e$-values~$(t,M_t)$, i.e., $[n]\times[0,1/\alpha] $ for a finite sample size~$n$&\ref{sect:DO_e} \\
              All-or-nothing bet  & Betting all the capital on a success in the next analysis &\ref{sect:DO_e}\\
                 Curtailment by bankruptcy  & Deciding to stop the trial because the $e$-value/capital reaches zero and betting cannot continue  &\ref{sect:DO_e}\\
                  Hopeless zone  & States in the~$e$-space  where it is no longer possible to hit the efficacy boundary before the horizon &\ref{sect:DO_e}\\
                   Almost hopeless zone  & States in the~$e$-space  where after an additional loss it is no longer possible to hit the efficacy boundary before the horizon &\ref{sect:DO_e}\\
                   P-max $e$-value   & The design-optimal $e$-value maximizing statistical power &\ref{sect:DO_e}\\
                     ESS-min $e$-value   & The design-optimal $e$-value minimizing the expected sample size &\ref{sect:DO_e}\\
                   $e$-value-based design   & 
                   Multi-stage single-arm binary data design with optimal futility and efficacy stopping rules based on $e$-values &\ref{sect:integrating_eval_design}\\ 
                   Fully sequential analysis in blocks   & 
                    Applying a fully sequential betting strategy for blocks of participants,
adhering to the arrival order &\ref{remark:n_interims}\\ 
		\hline
	\end{tabular}}
\end{table}

\end{document}